\documentstyle[aps,psfig]{revtex}
\newcommand{\s}{\mathrm}
\newcommand{\nunu}{\nonumber}
\newcommand{\bd}{\mathbf}
\newcommand{\ra}{\rightarrow}
\newcommand{\mn}{\mu \nu}
\newcommand{\be}{\begin{equation}}
\newcommand{\ee}{\end{equation}}

\newcommand{\ba}{\begin{eqnarray}}
\newcommand{\ea}{\end{eqnarray}}
\newcommand{\bef}{\begin{figure}}
\newcommand{\eef}{\end{figure}}

\newcommand{\rpp}{\rho \pi \pi}

\begin{document}


\begin{center}
{\Large{Electromagnetic Radiation from Hot and Dense Hadronic Matter}\\
}
\vskip .2in
Pradip Roy, Sourav Sarkar and Jan-e Alam\\
{\it Variable Energy Cyclotron Centre,
     1/AF Bidhan Nagar, Calcutta 700 064
     India}\\

Bikash Sinha\\

{\it Variable Energy Cyclotron Centre,
     1/AF Bidhan Nagar, Calcutta 700 064
     India}\\
{\it Saha Institute of Nuclear Physics,
           1/AF Bidhan Nagar, Calcutta 700 064
           India}\\
\end{center}
\vskip .2in

\begin{abstract}
The modifications of hadronic masses and decay widths at finite temperature 
and baryon density are investigated using a phenomenological model of 
hadronic interactions in the Relativistic Hartree Approximation.
We consider an exhaustive set of hadronic reactions 
and vector meson decays to estimate the photon emission from hot and dense 
hadronic matter.  The reduction in the vector meson masses and decay widths 
is seen to cause an enhancement in the photon production. 
It is observed that the effect of $\rho$-decay width 
on photon spectra is negligible.
The effects on dilepton production from pion annihilation are also indicated.
\end{abstract}



\addtolength{\baselineskip}{0.4\baselineskip}

\section*{I. Introduction}
 Numerical simulations of QCD (Quantum Chromodynamics) equation of state 
on the lattice predict that at very high density and/or temperature 
hadronic matter undergoes a phase transition to Quark Gluon Plasma 
(QGP)~\cite{ukawa,hwa}. One expects that ultrarelativistic heavy ion 
collisions might create conditions conducive for the formation and study 
of QGP. Various model calculations have been
performed to look for observable signatures of this state of matter.
However, among various signatures of QGP, photons and
dileptons are known to be advantageous as these signals participate only in 
electromagnetic interactions, and therefore, retain their energy and momentum
almost undistorted. They thus carry the information of the reaction zone
(from where they were produced) rather more effectively unlike the
hadrons which are strongly interacting and thus the signals get distorted
by final state interaction.
The disadvantage with photons is the substantial background from 
various processes (thermal and non-thermal)~\cite{janepr}. Among these, 
the contribution from hard QCD processes is well understood in the 
framework of perturbative QCD and the yield from hadronic decays 
e. g. $\pi^0\,\ra\,\gamma\,\gamma$ can be accounted for by invariant mass 
analysis. However, photons from the thermalised hadronic gas pose a more
difficult task to disentangle. Therefore it is very important to estimate
photons from hot and dense hadronic gas along with the possible modification
of the hadronic properties.   

    In ultrarelativistic heavy ion collisions hadronic matter is formed
after a phase transition from QGP. Even if such a
phase transition does not occur, realisation of hadronic matter at
high temperature ($\sim$ 150 -- 200 MeV) and/or baryon density ( a few
times normal nuclear matter density) is inevitable. As a result
the study of hadronic interactions at high temperature and density
assumes great significance. Such a study is also important in order to
understand the properties of compact stellar objects such as neutron stars 
where densities $\sim$ five to ten times normal nuclear matter
density are likely to be encountered. There are several other aspects
where medium effects may play an important role. For example spontaneously
broken chiral symmetry of the normal hadronic world is expected to be restored 
at high temperature and density~\cite{pisarski} and this will be 
reflected in the thermal shift of the hadron masses.
These modifications can be studied by analysing photon, dilepton as well
as hadronic spectra.

     Progress in our understanding of hot and dense hadronic matter has
been retarded since the underlying theory for strong interaction, QCD,
is nonperturbative at low energy regime. Because of this severe 
constraint considerable amount of work has been done on model 
building (see e.g. Ref.~\cite{meissner}) 
in order to study the low energy hadronic states. 

Various investigations have addressed the issue of temperature and density
dependence of hadronic masses  within different models over the
past several years. Hatsuda and collaborators~\cite{furn}  and 
Brown~\cite{brown} have  used the QCD sum rules  
at finite temperature and density to study the  
effective masses of the hadrons.
Brown and Rho~\cite{rho}  also argued that requiring chiral symmetry 
(in particular the QCD
trace  anomaly)  yield  an  approximate  scaling relation between
various effective hadronic masses,
$m_N^\ast/m_N\sim m_{\sigma}^\ast/m_{\sigma}\sim
m_{\rho,\omega}^\ast/m_{\rho,\omega}
\sim f_\pi^\ast/f_\pi$
which  implies,  that  all
hadronic   masses   decrease  with temperature. 
The nonlinear sigma model, claimed to be the closest
low energy description of QCD~\cite{meissner} however, shows the opposite
trend, {\it {i.e.}} $m_\rho^\ast$ increasing with temperature~\cite{abhijit}.
Thus,  there  exists  a lot of controversy in the literature about
the  finite  temperature   properties   of   hadrons. 

The nucleon-nucleon interaction is well described by semi-phenomenological
one-boson-exchange models. Inclusion of heavier mesons e. g. rho, omega and
other multipion resonant states in the one-pion exchange model gives a good
description of low energy $NN$ scattering data. However, in symmetric
nuclear matter the interaction mediated by pion and rho exchange 
averages out to zero. Therefore, many of the observed properties of
hadronic interactions at low energy can be understood by considering 
only omega and sigma meson exchange~\cite{vol16}.

   The change in the masses and decay widths of vector mesons propagating in
a medium occurs due to its interaction with the real and virtual
excitations in the medium. In Mean Field Theory (MFT) the condensed scalar
field is responsible for the modification of the nucleon mass. In other
words the nucleon mass changes due to the contribution from scalar
tadpole diagrams in the nucleon self energy~\cite{vol16,chin}.
The vector meson mass gets shifted due to the 
decrease of the nucleon mass which appears through thermal loops in the
vector meson self energy. The response of the nuclear system to
the external probe is characterised by the imaginary part of the vector
meson self energy. The interaction of the vector meson with the real particles
(on-shell) present in the medium brings in a small change in the 
mass of the vector meson but the net reduction in the vector meson mass
can be attributed to its interaction with the nucleons in the Dirac sea. 

In an earlier calculation~\cite{npa1} we have shown 
the effects of temperature modified masses and widths on the photon spectra. 
This has relevance to the conditions likely to be achieved at RHIC and LHC 
energies. However, due to partial stopping of baryonic matter at the SPS 
energies, the central region of heavy ion collision
can have non-zero baryon density~\cite{qm96}.
Hence it is necessary to include finite baryon density effects in the 
hadronic properties as well. 

    In the present calculation we have studied the medium modification of 
hadron properties both at finite temperature and baryon density. In addition 
to this we have also included the effects of vacuum fluctuations in the
nucleon self energy. The photon emission rates are then
estimated with medium modified mass and decay width of hadrons.
Possible effects on the dilepton emission rate have also been
illustrated.

  We organise the paper as follows. In Section II we calculate the temperature
and density dependent properties of vector mesons within the framework of
Quantum Hadrodynamics (QHD).
In Section III we discuss photon  production rates from
hot and dense matter. Section IV is devoted to discuss the results of 
our calculations. In Section V we present a summary and discussions.

\section*{ II. Medium Effects}
At non-zero temperature and density the pole of the propagator gets
shifted due to interactions with real and virtual excitations present
in the system. Such a modification can be studied through the
Dyson - Schwinger equation. 

\subsection*{ IIa. Nucleons}
In the Relativistic Hartree Approximation (RHA) the full nucleon 
propagator is given by,
\be
G^H(k) = G^0(k) + G^0(k)\Sigma^H(k)G^H(k)
\label{rhadyson}
\ee 
where $\Sigma^H(k)$ is the nucleon self energy which contains contributions
from both scalar ($\Sigma_s$) and vector ($\Sigma_v^{\mu}$) tadpole
diagrams~\cite{vol16,chin} and is given by

\be
\Sigma^H = \Sigma_s^H - \gamma^{\mu}\Sigma_{\mu v}^H
\label{sigma}
\ee
where,
\be
\Sigma_s^H = i\frac{g_s^2}{m_s^2}\,\int\,\frac{d^4q}{(2\pi)^4}\,
{\s {Tr}}[G^H(q)] + \Sigma_s^{\s{CTC}}
\ee 
and,
\be
\Sigma_{\mu v}^H = i\frac{g_v^2}{m_v^2}\,\int\,\frac{d^4q}{(2\pi)^4}\,
{\s {Tr}}[\gamma_{\mu}\,G^H(q)]
\ee
Here, $m_s\,\,(m_v)$ is the mass of the neutral scalar (vector) meson
, and, the nucleon interacts via the exchange of
scalar (vector) meson with coupling constant $g_s\,\,(g_v)$.
$ \Sigma_s^{\s{CTC}}$ is the counter term contribution required to
subtract the divergences in the scalar self energy. Since the
vector self energy is finite, such a counter term
is not required~\cite{vol16,chin}. The pole structure of the
full nucleon propagator in RHA resembles that of the non-interacting
propagator with shifted mass and four-momentum. The full nucleon
propagator consists of a medium and a vacuum part.
Mean Field Theory is reproduced if one considers only the
medium contribution. However, the inclusion of the vacuum
part  results in divergences. One then introduces counterterms
($\Sigma_s^{\s{CTC}}$) in order to subtract out the divergences. 
These constitute the vacuum fluctuation corrections to MFT. Consequently,
the effective nucleon mass reads,
\ba
\Sigma_s^H & = &M^{\ast} - M\nonumber\\
           & = & -\frac{g_s^2}{m_s^2}\,\frac{4}{(2\pi)^3}\int\,
d^3{\bd k}\,\frac{M^{\ast}}{E^\ast}\,
[n_B(\mu^{\ast},T)+{\bar n}_ B(\mu^{\ast},T)]\nonumber\\
           & + & \frac{g_s^2}{m_s^2}\,\frac{1}{\pi^2}\left[M^{\ast 3}
{\s {ln}}\left(\frac{M^{\ast}}{M}\right)-M^2(M^{\ast}-M)\right.\nonumber\\
           & - & \left.\frac{5}{2}M(M^{\ast}-M)^2-\frac{11}{6}(M^{\ast}-M)^3
\right]
\label{nmass}
\ea

where,
\ba
n_B(\mu^{\ast},T)& = &\frac{1}{{\s {exp}}[(E^{\ast}-\mu^{\ast})/T]+1}\nonumber\\
{\bar n}_B(\mu^{\ast},T)& = &\frac{1}{{\s {exp}}[(E^{\ast}+\mu^{\ast})/T]+1}
\nonumber\\
E^{\ast} & = & \sqrt{({\bd k}^2+M^{\ast 2})}\nonumber\\
\mu^{\ast} & = & \mu - \left(\frac{g_v^2}{m_v^2}\right)\rho\nonumber
\ea
Here, $\rho$ is the baryon density of the medium and is given by
\be
\rho = \frac{4}{(2\pi)^3}\int\,d^3{\bd k}\,
[n_B(\mu^{\ast},T)-{\bar n}_ B(\mu^{\ast},T)]
\label{bden}
\ee

\subsection*{ IIb. Vector Mesons}

The effective propagator for the vector boson is given by,

\be
D_{\mn} = -\,\frac{A_{\mn}}{k^2-m_V^2+\Pi_{T}}
-\,\frac{B_{\mn}}{k^2-m_V^2+\Pi_{L}} + \frac{k_{\mu}k_{\nu}}{k^2\,m_V^2},
\label{deff}
\ee
where $A_{\mn}$ and $B_{\mn}$ are the projection operators~\cite{npa1},
$m_V$ is the free mass of the vector meson and
\be
\Pi_{T(L)}=\Pi_{T(L)}^D+\Pi^F.
\ee
The transverse (longitudinal) component of the self energy,
$\Pi_{T(L)}^D$ contains both finite temperature and density effects
and $\Pi^F$ represents the contribution from the Dirac sea
with modified nucleon mass. We use the following interaction Lagrangian 
to evaluate the rho and omega self energies:
\be
{\cal L}_{VNN} = g_{VNN}\,\left({\bar N}\gamma_{\mu}
\tau^a N{V}_{a}^{\mu} - \frac{\kappa_V}{2M}{\bar N}
\sigma_{\mu \nu}\tau^a N\partial^{\nu}V_{a}^{\mu}\right),
\label{lag1}
\ee
where $V_a^{\mu} = \{\omega^{\mu},{\vec {\rho}}^{\mu}\}$,
$M$ is the free nucleon mass, $N$ is the nucleon field
and $\tau_a=\{1,{\vec {\tau}}\}$.

The real part of the self energy is responsible for mass shifting and
the imaginary part gives the decay width of the vector meson in the medium.
The physical mass of the vector meson can be obtained from
the pole position of the propagator in the limit $|\bd k|\,\rightarrow 0$
{\it i. e.} from the solution of the longitudinal or transverse
dispersion relation in the rest frame of the vector meson
as given in appendix A. 
The rho decay width is calculated from the $\rho\,\pi\,\pi$
loop by using Landau - Cutkosky cutting rules at finite 
temperature, the details being presented in appendix A.

\subsection*{IIc. Spectral Function for Vector Mesons}
The density of a stable hadron of mass $m$ in a thermal bath is completely
determined by the temperature, chemical potential and 
the statistics obeyed by the species  through the following 
equation
\be
\frac{dN}{d^3x\,d^3k\,ds}=\frac{g}{(2\pi)^3}\frac{1}{\exp(k_0-\mu)/T\pm\,1}\,
\delta(s-m^2)
\label{stable}
\ee
where $g$ is the statistical degeneracy and $k_0=\sqrt{\vec k^2\,+s}$ 
is the energy of the particle in the rest frame of the thermal
bath.  The question one would like to ask is - how 
will eq.~(\ref{stable}) be modified if the 
particle decays within the thermal bath? This is a relevant question 
for $\rho$ and $\omega$ mesons as their life times ( 1.3 fm/c and 23.5 fm/c 
respectively in vacuum) are comparable to the strong interaction 
time scale and therefore they can decay within the thermal system
formed after ultra-relativistic heavy ion collisions. This problem 
has been addressed by Weldon~\cite{weldonann} through 
the generalisation of Breit-Wigner formula at finite 
temperature and density and its consequences on the dileptonic decay of
vector mesons have been studied in ~\cite{prc99}. The distribution 
of an unstable particle in a thermal bath is given by~\cite{weldonann}
\be
\frac{dN}{d^3x\,d^3k\,ds}=\frac{g}{(2\pi)^3}
\frac{1}{\exp(k_0-\mu)/T\pm\,1}\,P(s)
\label{unstable}
\ee
where $P(s)$ is called the spectral function~\cite{bellac,abrikosov,zubarev} 
can be calculated from the effective thermal propagator. It is  given by
\be
P(s)=\left[\frac{1}{\pi}
\frac{{\s {Im}}\Pi}{(s-m_V^2+{\s {Re}}\Pi)^2+({\s {Im}}\Pi)^2}\right]
\ee
Equation~(\ref{unstable}) indicates that to obtain 
realistic results for the photon production through a 
reaction involving unstable particle in the external 
line the finite width of the particle should be taken
into account by introducing the spectral representation
of the corresponding particle and integrating over $s$~\cite{rapp}.
This is done in our calculation for the unstable vector mesons
appearing in the external line in reactions for photon production.
In case the unstable particle appears in the internal
line the finite width of the particle is taken into account
through effective propagators.
It is interesting to  note that the spectral function reduces to a 
Dirac delta function $\delta(s-m_V^2+{\s {Re}}\Pi)$
in the limit ${\s {Im}}\Pi\,\ra\,0$ i.e. when the particle is stable. 
In the calculation of the imaginary part of the self energy ${\s {Im}}\Pi$
of $\rho$ say, one must in principle, include all the processes 
which can create or annihilate a $\rho$ in the thermal bath. 
However, within the ambit of the model adopted in the present work
we have seen that the most dominant contribution to 
${\s {Im}}\Pi$ comes from the $\rho-\pi-\pi$ interaction
in the temperature and density range of our interest. 

\section*{III. Photon emission}
To evaluate the photon emission rate from a hadronic gas
we  model the system as consisting of $\pi$, $\rho$, $\omega$
and $\eta$. The relevant vertices for the reactions 
$\pi\,\pi\,\ra\,\rho\,\gamma$ and $\pi\,\rho\,\ra\,\pi\,\gamma$
and the decay $\rho\,\ra\,\pi\,\pi\,\gamma$
are obtained from the following Lagrangian:
\be
{\cal L} = -g_{\rho \pi \pi}{\vec {\rho}}^{\mu}\cdot
({\vec \pi}\times\partial_{\mu}{\vec \pi}) - eJ^{\mu}A_{\mu} + \frac{e}{2}
F^{\mu \nu}\,({\vec \rho}_{\mu}\,\times\,{\vec \rho}_{\nu})_3,
\label{photlag}
\ee
where $F_{\mu \nu} = \partial_{\mu}A_{\nu}-\partial_{\nu}A_{\mu}$, is the
Maxwell field tensor and $J^{\mu}$ is the hadronic part of the electromagnetic
current given by
\be
J^{\mu} = ({\vec \rho}_{\nu}\times{\vec B^{\nu \mu}})_3 + (
{\vec \pi}\times(\partial^{\mu}{\vec \pi}+g_{\rho \pi \pi}{\vec \pi}\times{\vec
\rho}^{\mu}))_3
\label{jmu}
\ee
with ${\vec B_{\mu \nu}} = \partial_{\mu}{\vec \rho}_{\nu}-\partial_{\nu}
{\vec \rho}_{\mu}-g_{\rho \pi \pi}(\vec \rho_{\mu}\times\vec \rho_{\nu})$.
The invariant amplitudes for all these reactions have been listed in 
the appendix of Ref.~\cite{npa1}.

For the sake of completeness we have also considered the photon 
production due to the reactions $\pi\,\eta\,\rightarrow\,\pi\,\gamma$, 
$\pi\,\pi\,\rightarrow\,\eta\,\gamma$ and the decay 
$\omega\,\ra\,\pi\,\gamma$ using the following interaction:
\be
{\cal L} =
\frac{g_{\rho \rho \eta}}{m_{\eta}}\,
\epsilon_{\mu \nu \alpha \beta}\partial^{\mu}{\rho}^{\nu}\partial^{\alpha}
\rho^{\beta}\eta
+\frac{g_{\omega \rho \pi}}{m_{\pi}}\,
\epsilon_{\mu \nu \alpha \beta}\partial^{\mu}{\omega}^{\nu}\partial^{\alpha}
\rho^{\beta}\pi
+\frac{em_{\rho}^2}{g_{\rho \pi \pi}}A_{\mu}\rho^{\mu}
\label{etaro}
\ee
The last term in the above Lagrangian is written down on the basis
of Vector Meson Dominance (VMD) ~\cite{sakurai}.
The invariant amplitudes for the reactions involving the $\eta$
meson are given in appendix B.

The emission rate of a photon of energy $E$ and momentum ${\bd p}$ 
from a thermal system at a temperature $T$ is given by
\ba
E\frac{dR}{d^3{\bd p}}&=&\frac{\cal N}{16(2\pi)^7E}\,\int_{(m_1+m_2)^2}^{\infty}
\,ds\,\int_{t_{\s {min}}}^{t_{\s {max}}}\,dt\,|{\cal M}|^2\,
\int\,dE_1\nonumber\\
&&\times\int\,dE_2\frac{f(E_1)\,f(E_2)\left[1+f(E_3)\right]}{\sqrt{aE_2^2+
2bE_2+c}}
\label{photrate}
\ea
where ${\cal M}$ is the invariant amplitude
for photon production and
\ba
a&=&-(s+t-m_2^2-m_3^2)^2\nonumber\\
b&=&E_1(s+t-m_2^2-m_3^2)(m_2^2-t)+E[(s+t-m_2^2-m_3^2)(s-m_1^2-m_2^2)\nonumber\\
&&-2m_1^2(m_2^2-t)]\nonumber\\
c&=&-E_1^2(m_2^2-t)^2-2E_1E[2m_2^2(s+t-m_2^2-m_3^2)-(m_2^2-t)(s-m_1^2-m_2^2)]
\nonumber\\
&&-E^2[(s-m_1^2-m_2^2)^2-4m_1^2m_2^2]-(s+t-m_2^2-m_3^2)(m_2^2-t)\nonumber\\
&&\times(s-m_1^2-m_2^2)
+m_2^2(s+t-m_2^2-m_3^2)^2+m_1^2(m_2^2-t)^2\nonumber\\
E_{1{\s {min}}}&=&\frac{(s+t-m_2^2-m_3^2)}{4E}+\frac{Em_1^2}{s+t-m_2^2-m_3^2}
\nonumber\\
E_{2{\s {min}}}&=&\frac{Em_2^2}{m_2^2-t}+\frac{m_2^2-t}{4E}\nonumber\\
E_{2{\s {max}}}&=&-\frac{b}{a}+\frac{\sqrt{b^2-ac}}{a}\nonumber.
\ea

\section*{IV. Results}
\subsection*{IVa. Hadronic Properties in the Medium}
The effective nucleon mass (which appears in the nucleon loop 
contribution to self energies of the rho and 
omega mesons) has been calculated, within the framework of 
the model defined above, in the RHA.  
The following values of the coupling constants
and masses~\cite{jean} have been used in our calculations: 
$\kappa_{\rho} = 6.1,~g_{\rho NN}^2 = 6.91, m_s$= 458 MeV, $m_{\rho} = 
770$ MeV, $M = 939$ MeV, $g_s^2 = 54.3$, $\kappa_{\omega} = 0$,
and $g _v^2 \equiv g_{\omega NN}^2 = 102$. In Fig.~(\ref{fig1}) we depict the
variation of nucleon mass with temperature for a set of
densities.  We observe that the nucleon mass falls steadily 
with density for a fixed temperature. However, the variation
with temperature for given values of baryon densities shows
interesting features. At zero baryon density the nucleon mass  
decreases monotonically as a function of temperature, but 
for finite densities it increases slightly before falling.
This trend is similar to that obtained by Li et al~\cite{li}, and
may be attributed to the modification of the Fermi-sea
at finite temperature and density. Our calculation shows a 35\%
reduction of the effective nucleon mass at $T$ = 160 MeV and
two times normal nuclear matter density 
compared to its free mass. In order to highlight the effect of vacuum 
fluctuation corrections we compare the MFT results with those
obtained using RHA. This is plotted in Fig.~(\ref{fig2}). We observe that
the effect of vacuum fluctuation is substantial for higher values of the
baryon density. The contribution of the antinucleons from the
Dirac sea is responsible for such an effect.     

\bef
\centerline{\psfig{figure=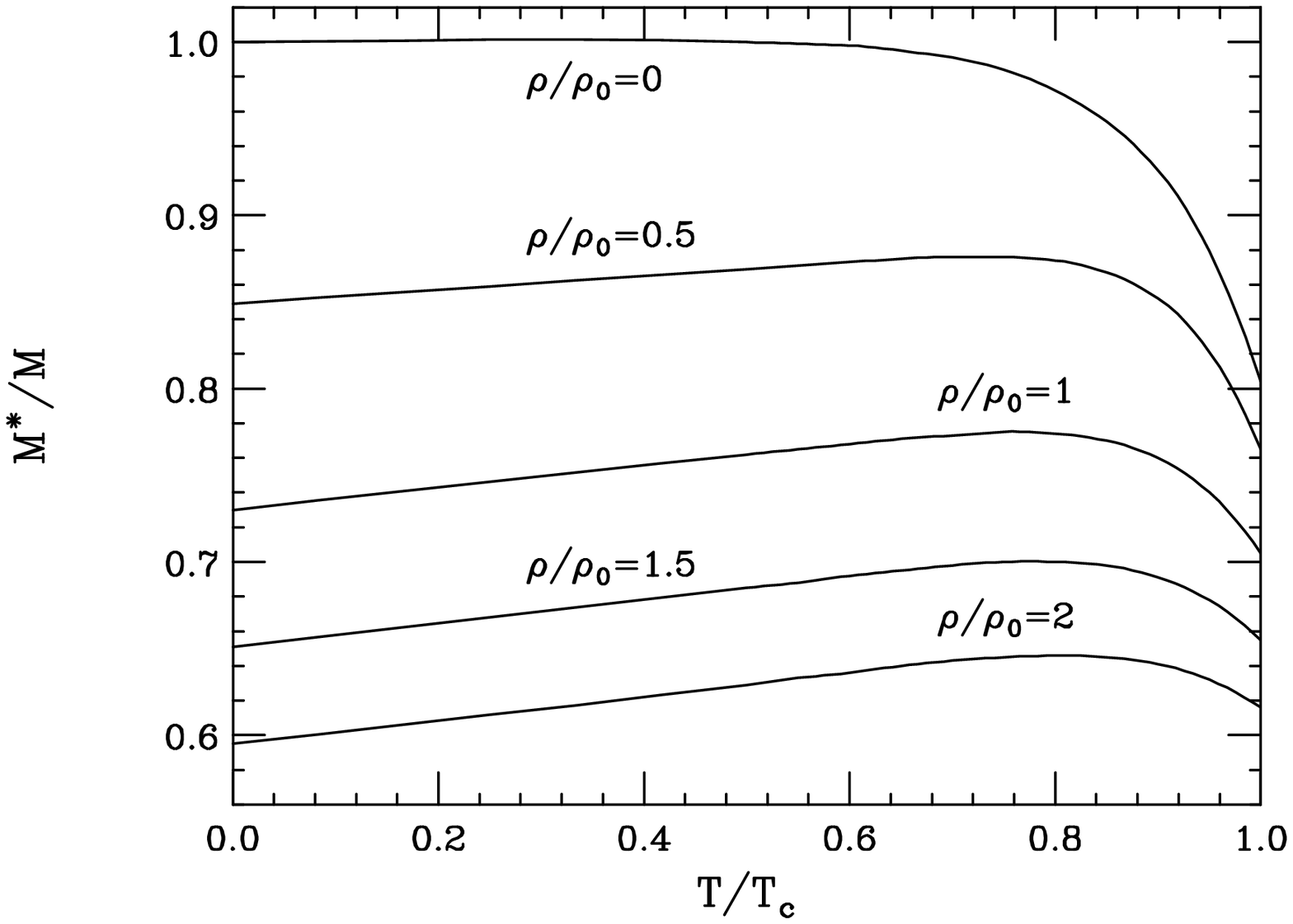,height=6cm,width=8cm}}
\caption{
Variation of nucleon mass with temperature 
for different values of baryon densities.
Here, $\rho_0$=0.1484 fm$^{-3}$ and $T_c$=200 MeV.
}
\label{fig1}
\eef

  We now focus our attention to the properties of vector mesons due
to $N-N$ interactions at finite temperature and baryon density.
The major contribution to the medium effects on the rho and omega mesons,
in this approach, arises from the nucleon-loop diagram.
For the dressing of internal lines in matter we restrict ourselves to 
Mean Field Theory (MFT) to avoid a plethora of diagrams and to maintain
internal consistency. 
Since the Walecka model does not have chiral symmetry,
it is rather difficult to predict anything reliable
on the pion mass in this model, especially in the MFT approximation
~\cite{vol16,kapusta}.
On the other hand, in the models with chiral symmetry
e.g. the Nambu-Jona-Lasinio model, and the linear sigma model
with nucleon, it is well-known that
the pion mass is almost unchanged in so far as one is in the 
Nambu-Goldstone phase. This is simply a consequence of the
Nambu-Goldstone theorem in medium~\cite{hatsuda1};
we thus adopt the approach of keeping the pion mass constant.
The longitudinal and transverse dispersion relations are
shown in Fig.~(\ref{fig3}) for two sets of coupling constants $(g_V, \kappa_V)$
$= (10.2, 0)$ and $(2.63, 6)$ relevant for omega and rho mesons respectively.
One observes a small difference between the two modes in case
of the omega meson. The vanishing of this difference for the rho meson
can be attributed to the smaller vector and larger tensor coupling constants
as compared to the omega meson.
In Fig.~(\ref{fig4}) the effective mass of the rho meson is plotted against
temperature for various values of baryon density. We observe that the
variation of the rho mass follows qualitatively the same trend as that of the 
nucleon. In this case, the rho mass decreases by 45\% at $T$= 160 MeV
and two times normal nuclear matter density compared to its free space value.
This is due to the fact that the large  decrease of the modified
Dirac sea contribution to the rho self energy dominates over the in-medium
contribution which is seen to increase with temperature.
It has been shown~\cite{npa1,gale} that the change in the rho mass due
to rho pion interaction is negligibly small at non-zero temperature
and zero baryon density. 
In a different model calculation it has been shown by 
Klingl et al~\cite{klingl} that up to 
to leading order in density the shift in the rho mass is very small.
Therefore the change in rho meson properties 
due to $\rho-\pi-\pi$ interaction is neglected here.
In another approach the in-medium 
modification in the spectral function of the rho meson was studied
~\cite{asakawa,friman,chanfray}
by including the medium effects on the $\rho-\pi-\pi$ vertex and 
the pion propagator in the delta-hole model at non-zero density. 
However, it is observed in these studies that the 
rho mass remains almost unchanged due to rho-pion interaction 
up to normal nuclear matter density.
Since in this work we 
restrict our calculations within the realm of MFT, {\it i.e} 
the internal nucleon loop in the rho and omega self energy 
are modified due to tadpole diagram only,
the inclusion of vertex corrections and modification of the pion 
propagator due to delta-hole excitation 
will take us beyond MFT and hence are not considered here.
However, if one includes the
delta-hole polarisation (and N-N polarisation also) effects on pions,
the rho mass may change from the values obtained here
due to $\rho\,-\,\pi$ interaction for $\rho>\rho_0$. 
We do not include the delta baryon in the present work.

\bef
\centerline{\psfig{figure=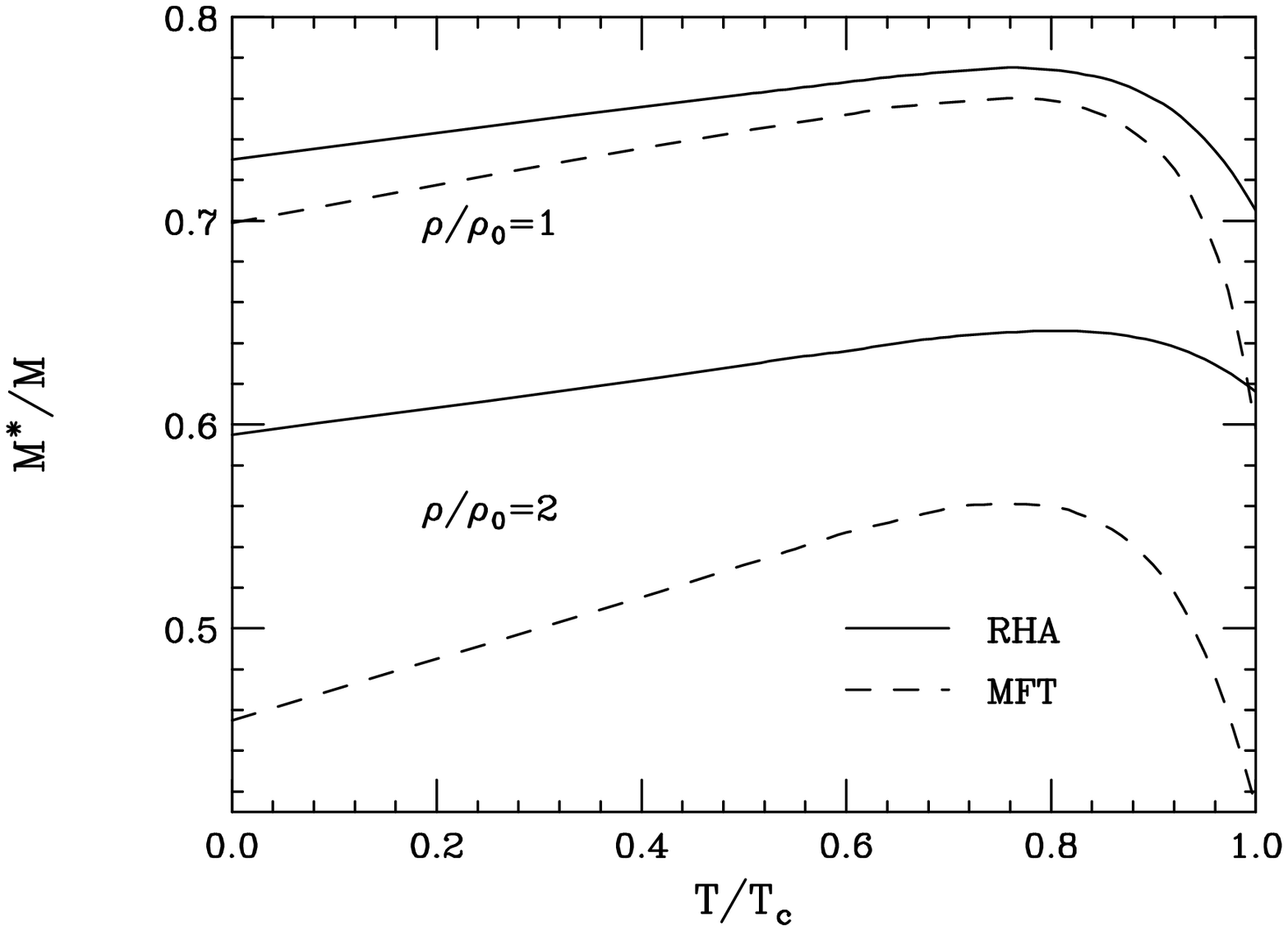,height=6cm,width=8cm}}
\caption{
Same as Fig.~\protect\ref{fig1} with (solid) and without( dashed)  
vacuum fluctuation corrections to MFT.
}
\label{fig2}
\eef


\bef
\centerline{\psfig{figure=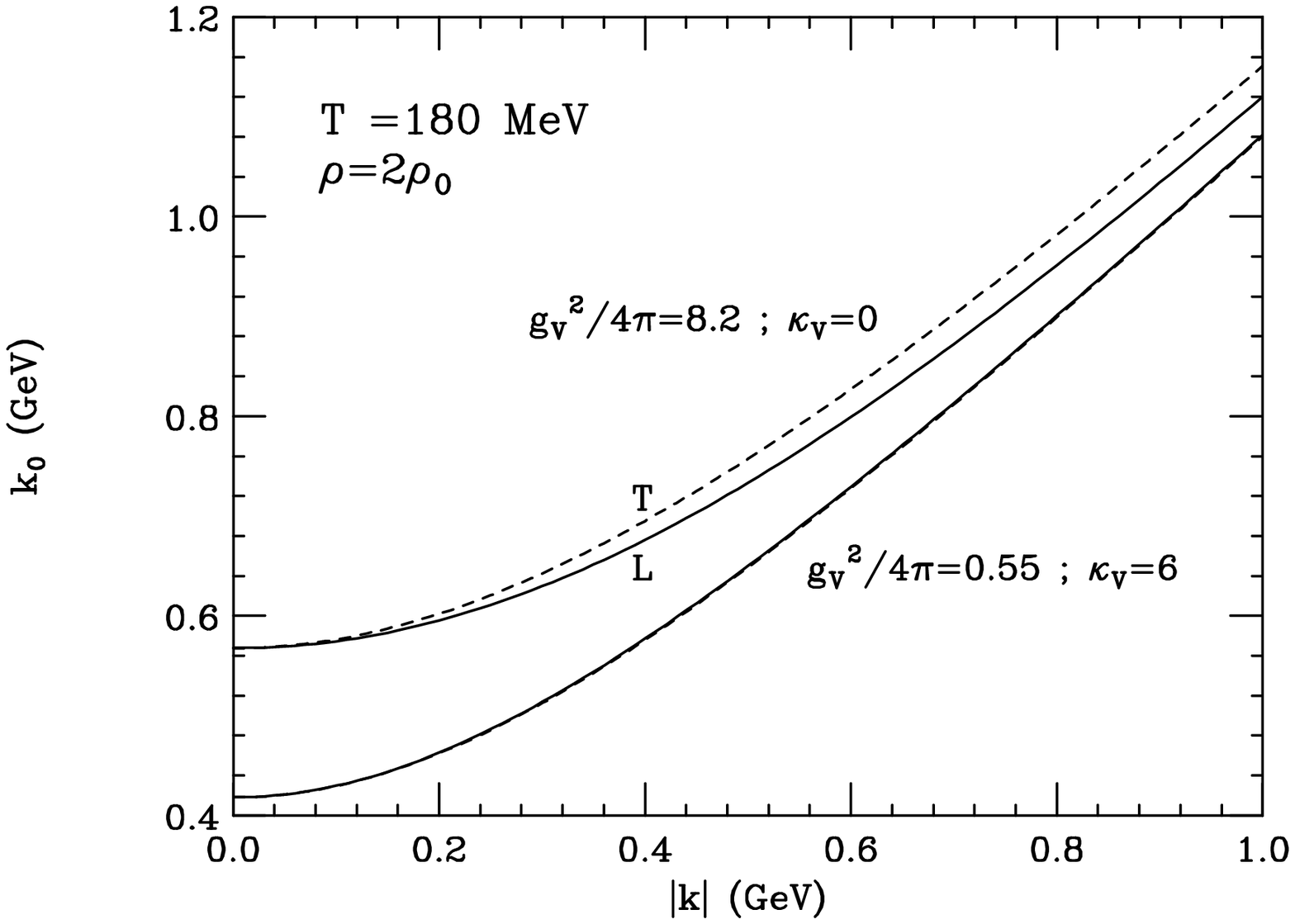,height=6cm,width=8cm}}
\caption{
Longitudinal and transverse dispersion relation for vector mesons.
}
\label{fig3}
\eef
 
We evaluate the effective omega mass with the
values of the  coupling constants mentioned above. We observe in 
Fig.~(\ref{fig5}) that the mass of the omega meson as a function of 
temperature follows qualitatively similar trend as that of the rho 
meson. The quantitative difference in the rho and omega meson masses 
is due to the different numerical values of the coupling constants e.g. 
the tensor interaction is absent in case of the omega meson. 

\bef
\centerline{\psfig{figure=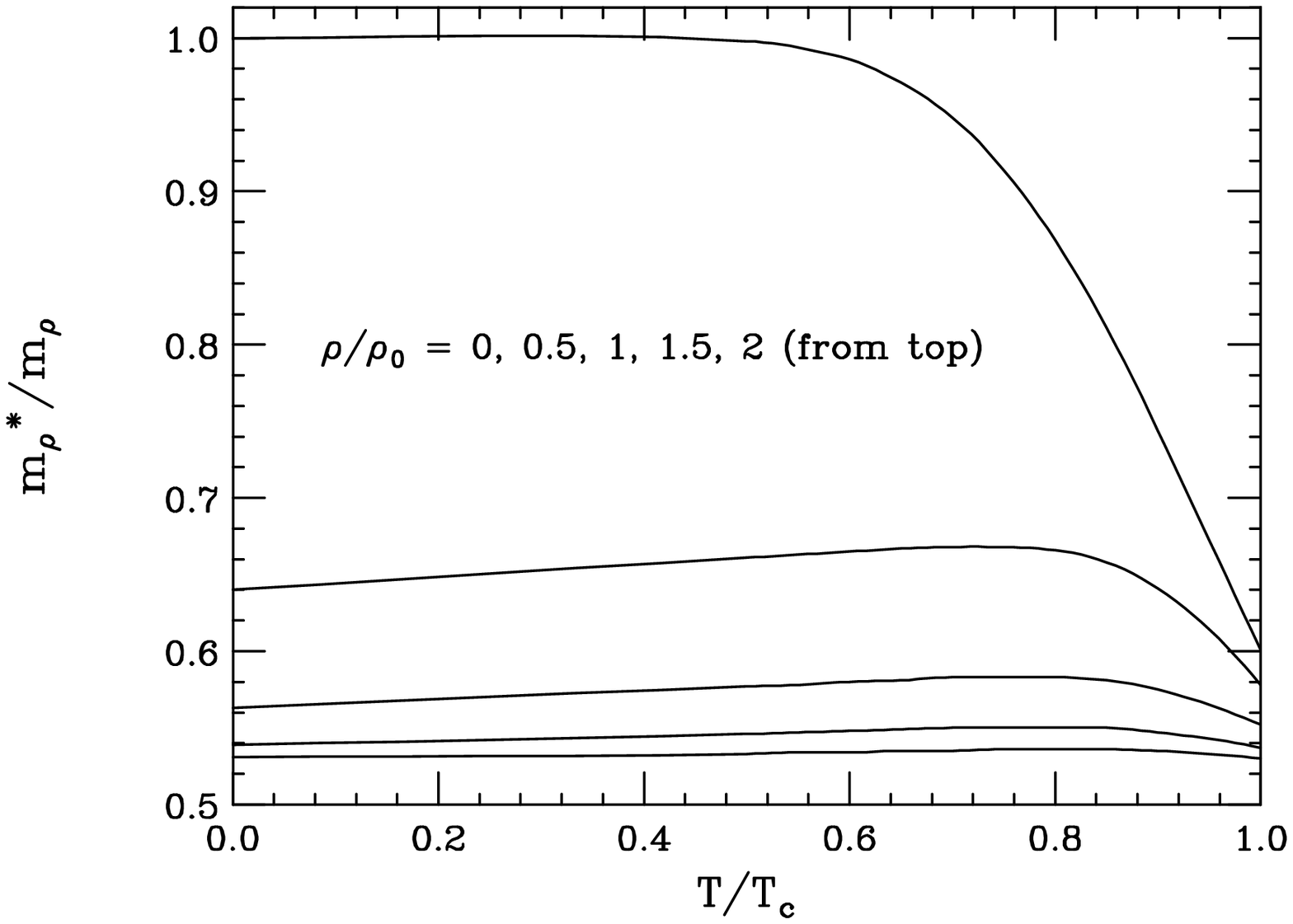,height=6cm,width=8cm}}
\caption{
Variation of rho mass with temperature for various baryon densities.
}
\label{fig4}
\eef

\bef
\centerline{\psfig{figure=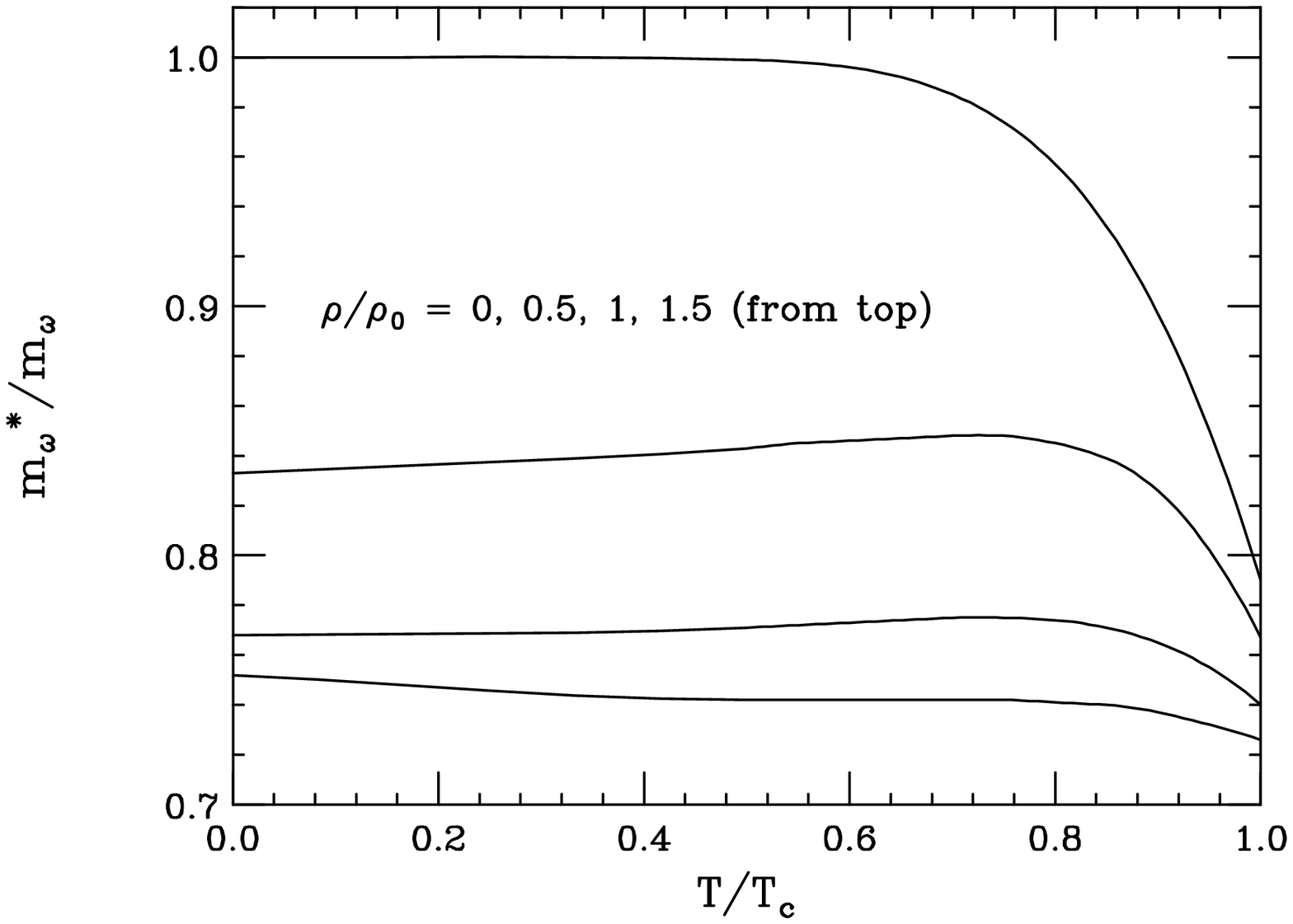,height=6cm,width=8cm}}
\caption{
Variation of omega mass with temperature for various baryon densities.
}
\label{fig5}
\eef

Now we demonstrate the in-medium effect on the decay width of
rho meson. This is shown in Fig.~(\ref{fig6}). In this case  
the observed enhancement of the decay width with temperature at
non-zero values of the baryon density is solely due to stimulated
emission of pions in the medium. This is just a manifestation of
the well known Bose enhancement (BE) effect ~\cite{npa1,abhee} which
is more clearly observed in Fig.~(\ref{fig7}). We will show 
in the next section that the BE effect plays an important
role in the dilepton spectra.

\bef
\centerline{\psfig{figure=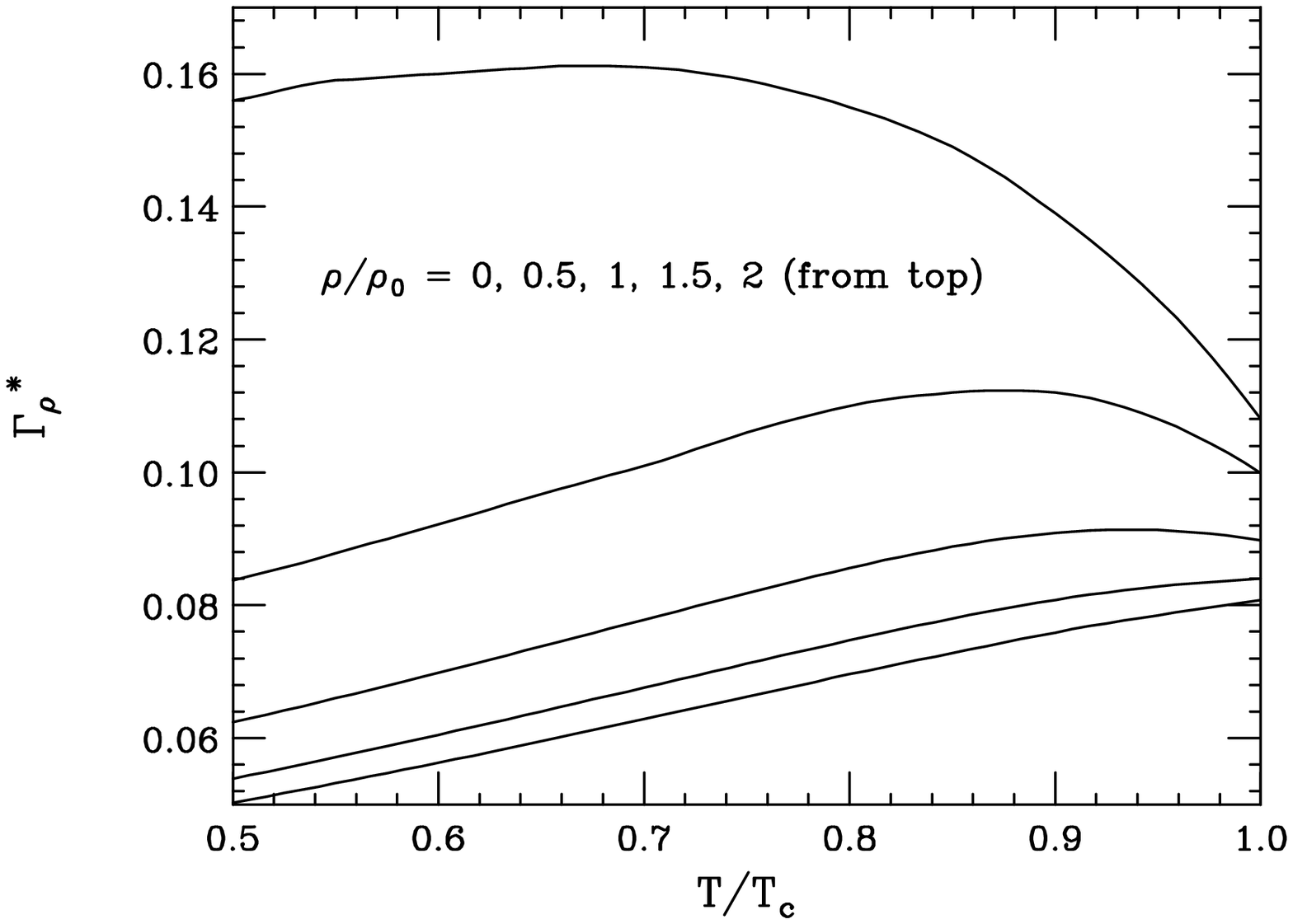,height=6cm,width=8cm}}
\caption{
The $\rho\,\ra\,\pi\,\pi$ decay width as a function of temperature
for different values of baryon densities.
}
\label{fig6}
\eef

\bef
\centerline{\psfig{figure=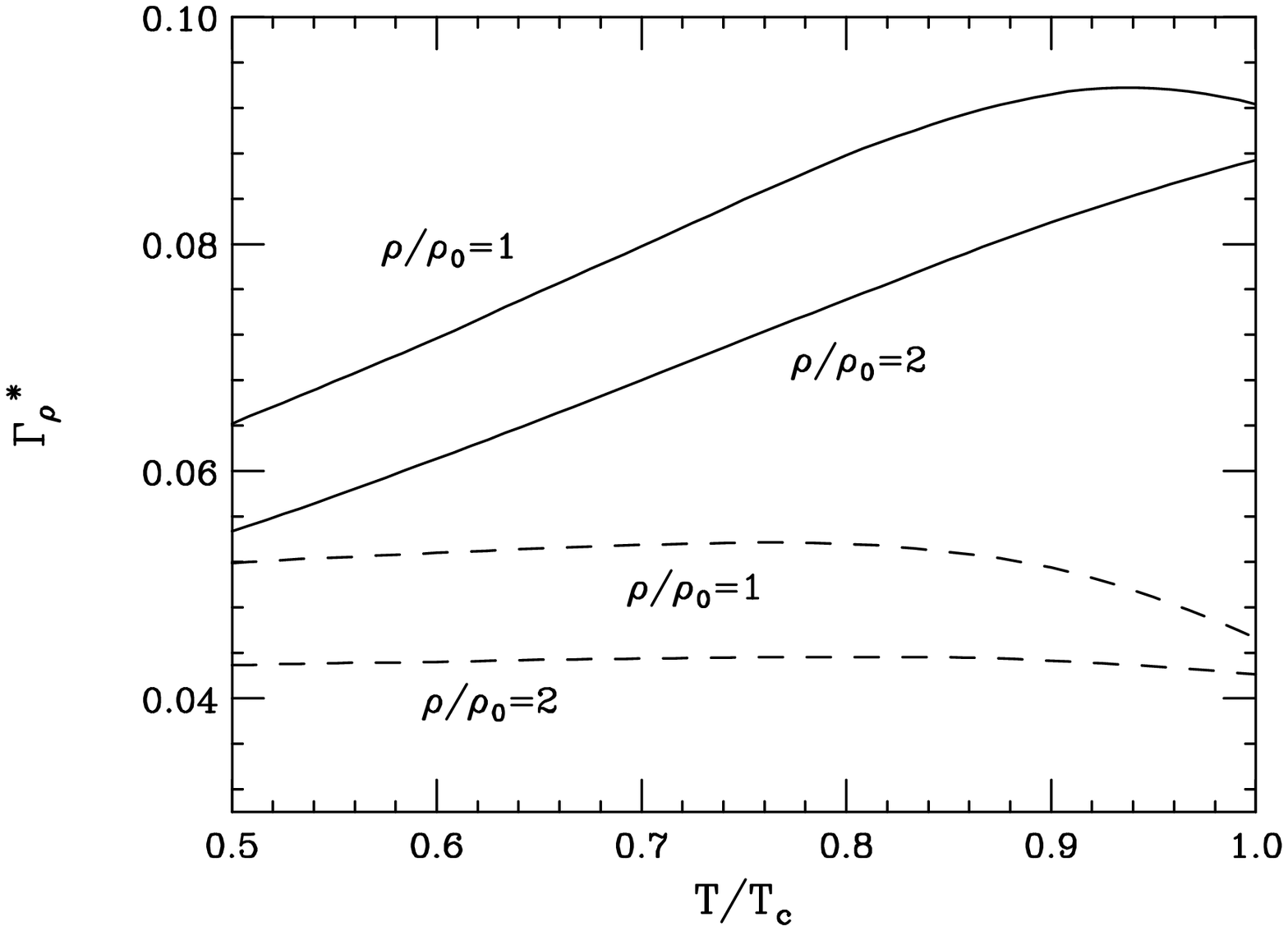,height=6cm,width=8cm}}
\caption{
Same as Fig.~\protect\ref{fig6} with (solid) and without (dashed)  
BE effect.
}
\label{fig7}
\eef

\subsection*{IVb. Photons from Hadronic Matter}

In the present section we evaluate the photon spectra from a hot and
dense hadronic medium. The medium effects enter through the 
polarisation functions in the vector meson propagator described by
eq.(\ref{deff}). 
It is well known~\cite{joe} that the reactions $\pi\,\rho\,\ra\, \pi\,\gamma$ , 
$\pi\,\pi\,\ra\, \rho\,\gamma$ , $\pi\,\pi\,\ra\, \eta\,\gamma$ , 
$\pi\,\eta\,\ra\, \pi\,\gamma$ , and the decays $\rho\,\ra\,\pi\,\pi\,\gamma$
and $\omega\,\ra\,\pi\,\gamma$ are the most important channels 
for photon production from hadronic matter in the
energy regime of our interest.
 
\bef
\centerline{\psfig{figure=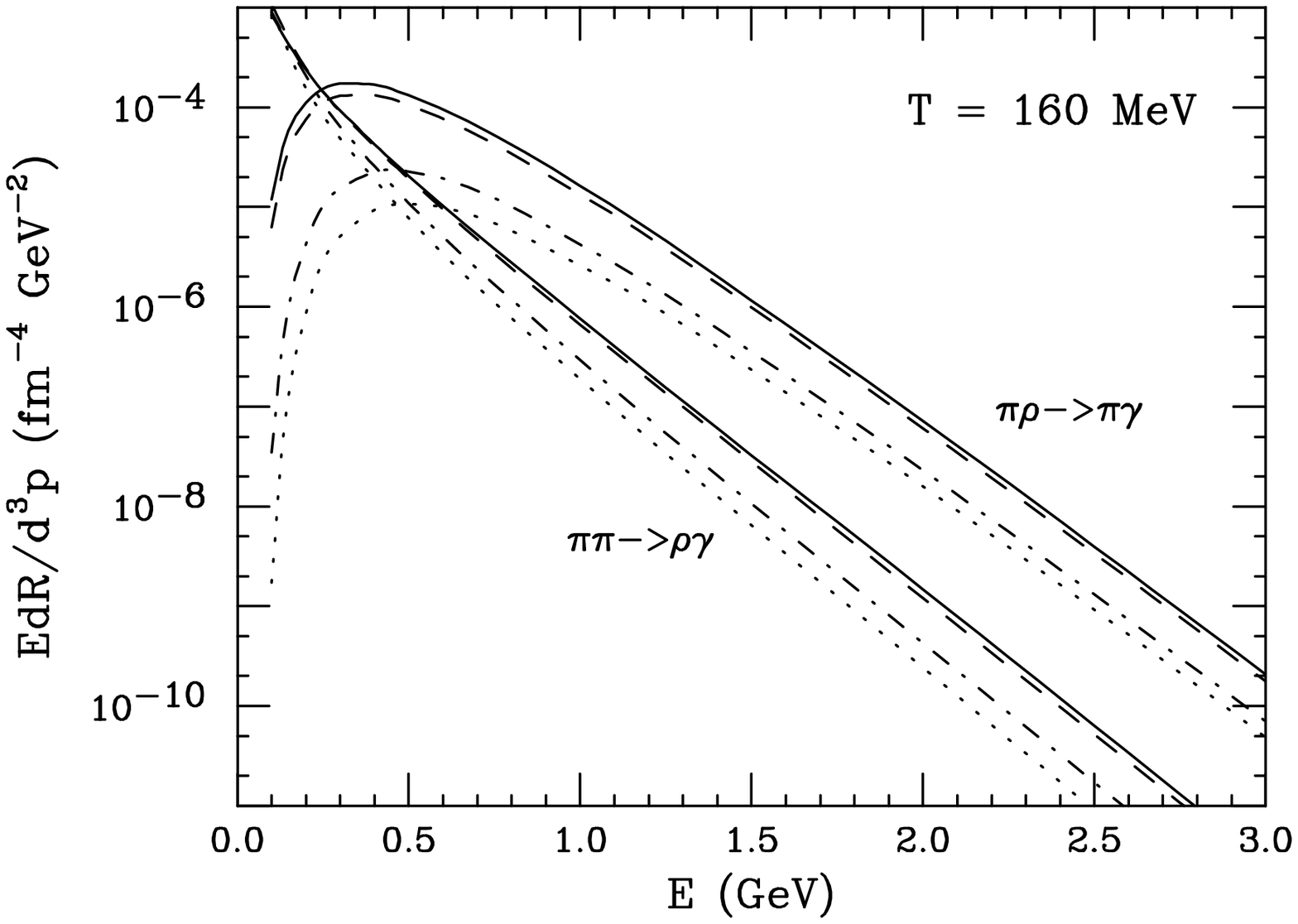,height=6cm,width=8cm}}
\caption{
Photon emission rates as a function of energy of the emitted photon
at $T$ = 160 MeV. Solid and dashed lines show
the results for $\rho/\rho_0$ = 2 and 1 respectively.
The dotdashed line represents only finite temperature effects and dotted
line shows the result when no medium effect is considered.
}
\label{fig8}
\eef

\bef
\centerline{\psfig{figure=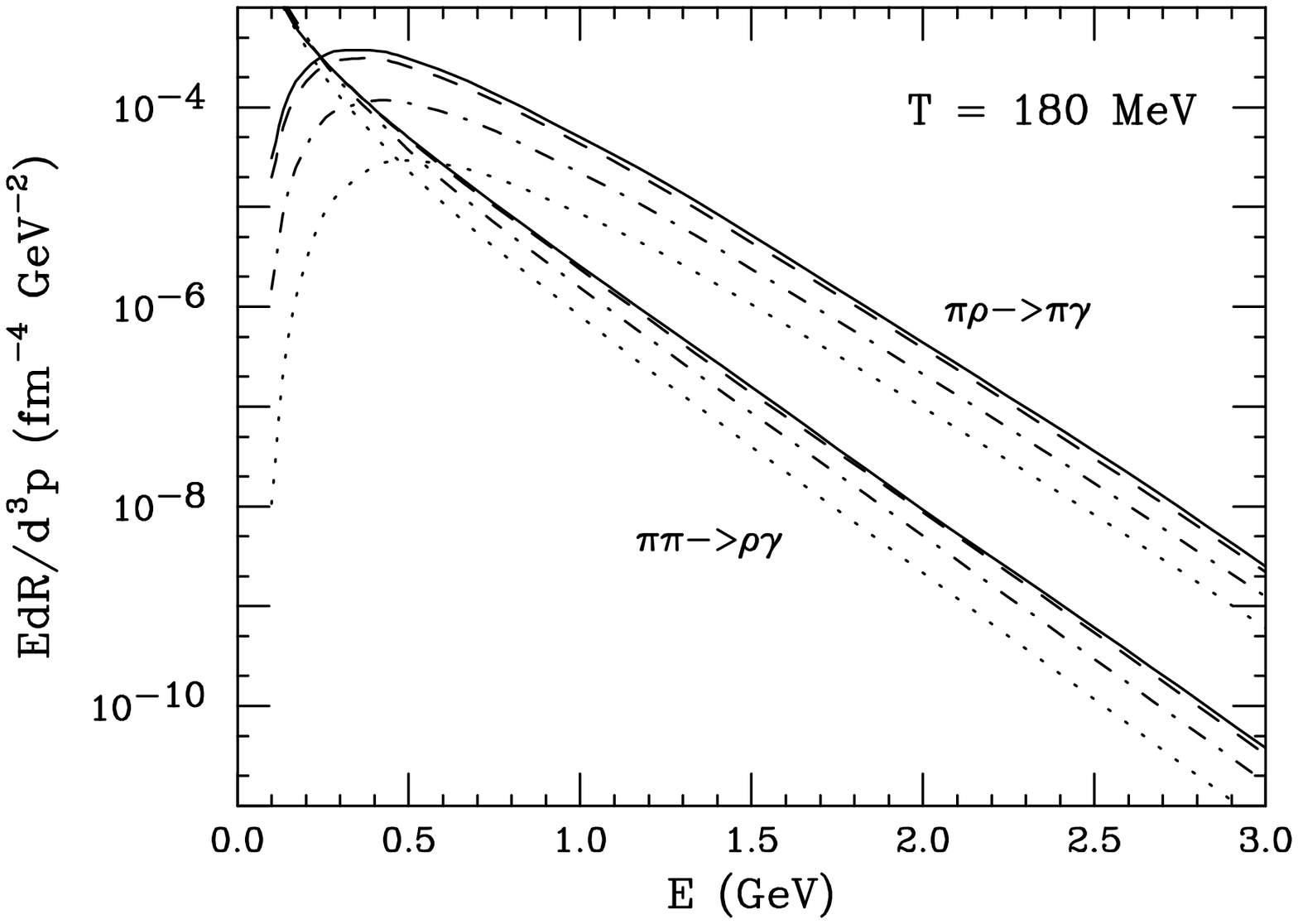,height=6cm,width=8cm}}
\caption{
Same as Fig.~\protect\ref{fig8} at $T$ = 180 MeV. 
}
\label{fig9}
\eef

We start with the reactions $\pi\,\rho\,\ra\, \pi\,\gamma$  and
$\pi\,\pi\,\ra\, \rho\,\gamma$. Fig.~(\ref{fig8}) shows the photon
yields from these reactions as a function of energy at $T$=160 MeV
for various values of the baryon density.
For the channel $\pi\,\pi\,\ra \,\rho\,\gamma$, we observe that
at two times normal nuclear density the emission rate of photons 
in the energy range 1-3 GeV increases by a factor of $\sim$ 6 compared
to its value when no medium effect is taken into account.  
We can understand this enhancement from a simple kinematical
argument. For a given center of mass energy of the $\pi-\pi$ system,
the energy of a photon can be written as $E_{\gamma} = (\sqrt s -m_{\rho}^2
/\sqrt s)/2$. One can immediately see from here that the decrease in rho mass
enhances the possibility of getting a photon of higher energy.
Let us now turn to the reaction $\pi\,\rho\,\ra\, \pi\,\gamma$.
In this case also we observe a similar enhancement in the photon yield
at higher energies.
However, at lower photon energies ($<$ 1 GeV) the enhancement
is by a few orders of magnitude. The rho meson 
in the incident channel in this case appears as a massless boson
($\gamma$) in the exit channel making available the rest mass energy
of the rho to the kinetic energy of the emitted photon.
In Fig.~(\ref{fig9}), the same is shown at a temperature $T$ = 180 MeV.
The results are qualitatively similar to the previous case with
an increase in the absolute value of the photon yield.

\bef
\centerline{\psfig{figure=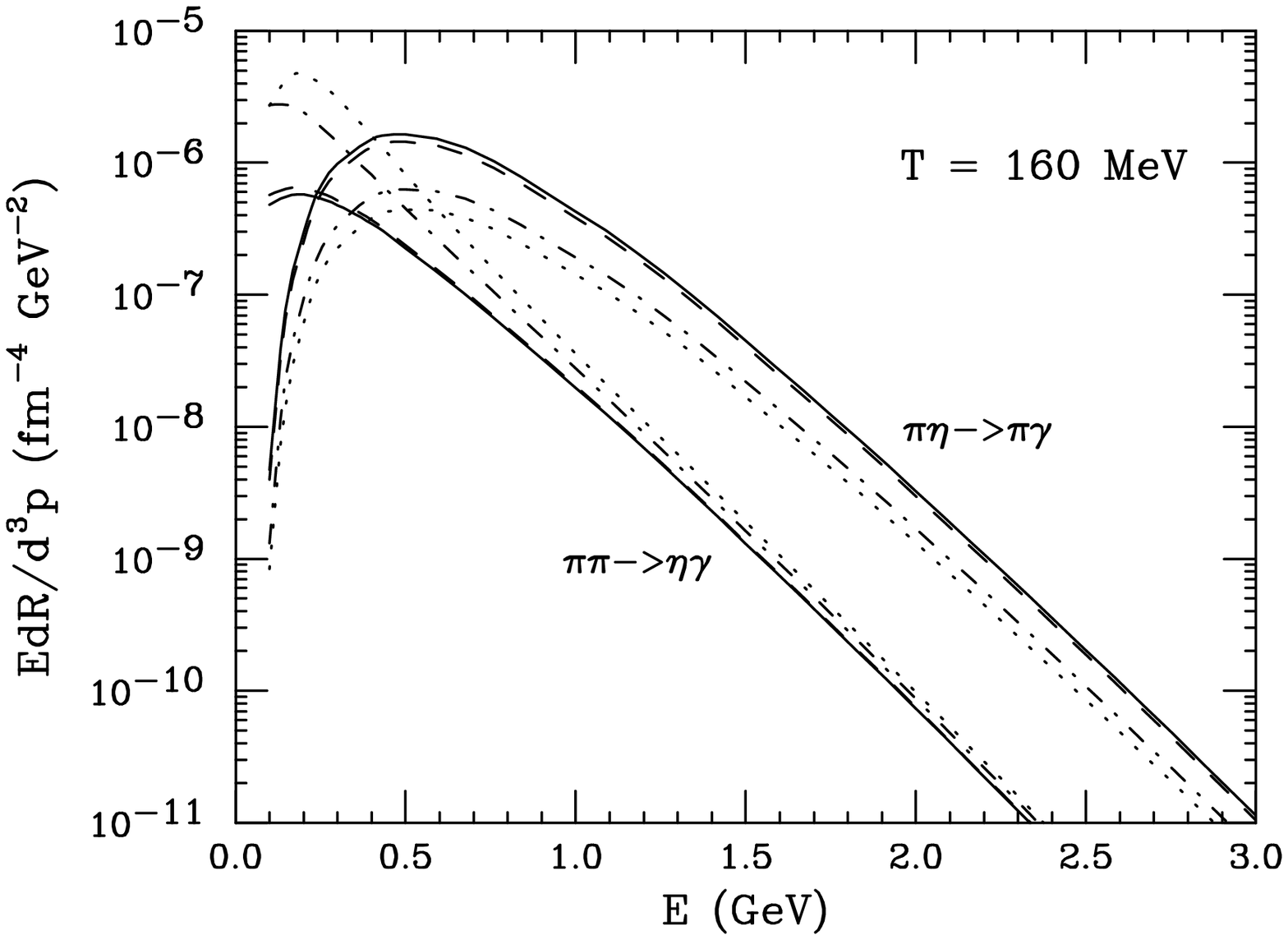,height=6cm,width=8cm}}
\caption{
Same as Fig.~\protect\ref{fig8} for the reactions $\pi\,\eta\,\ra\,
\pi\,\gamma$ and $\pi\,\pi\,\ra\,\eta\,\gamma$. 
}
\label{fig10}
\eef

\bef
\centerline{\psfig{figure=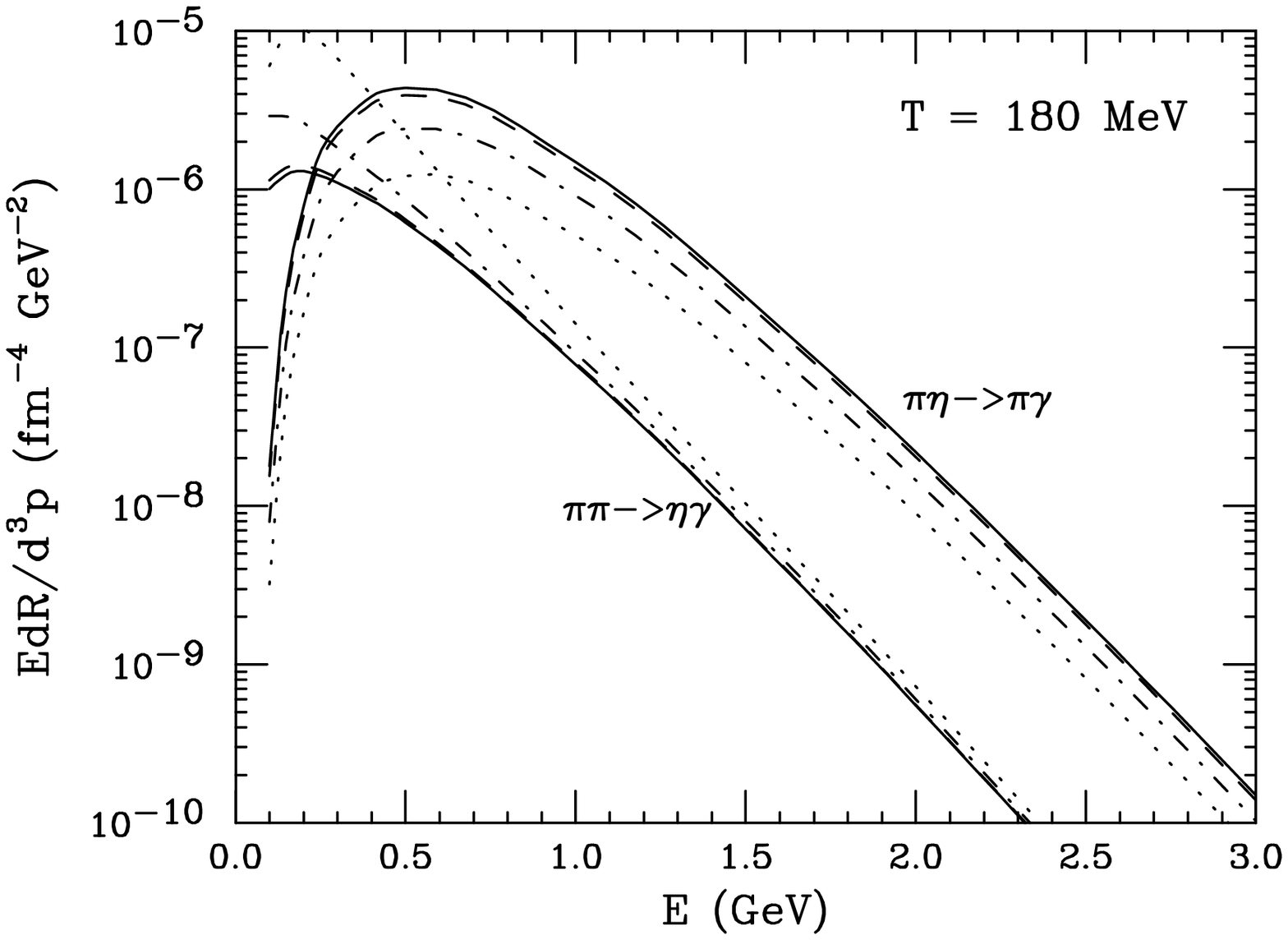,height=6cm,width=8cm}}
\caption{
Same as Fig.~\protect\ref{fig10} at $T$ = 180 MeV. 
}
\label{fig11}
\eef

We next consider the reactions $\pi\,\pi\,\ra\, \eta\,\gamma$, and 
$\pi\,\eta\,\ra\, \pi\,\gamma$. In Fig.~(\ref{fig10}) we plot
the photon emission rates from these reactions at $T$ = 160 MeV
for different baryon densities. It is interesting to note that
with increasing values of the baryon density the yield from
the first reaction decreases whereas the opposite occurs for the second
reaction. Such a behaviour naturally follows from the rho mass (effective)
dependence of the invariant amplitudes (see appendix B). The same features 
are also observed at $T$ = 180 MeV as shown in Fig.~(\ref{fig11}).

\bef
\centerline{\psfig{figure=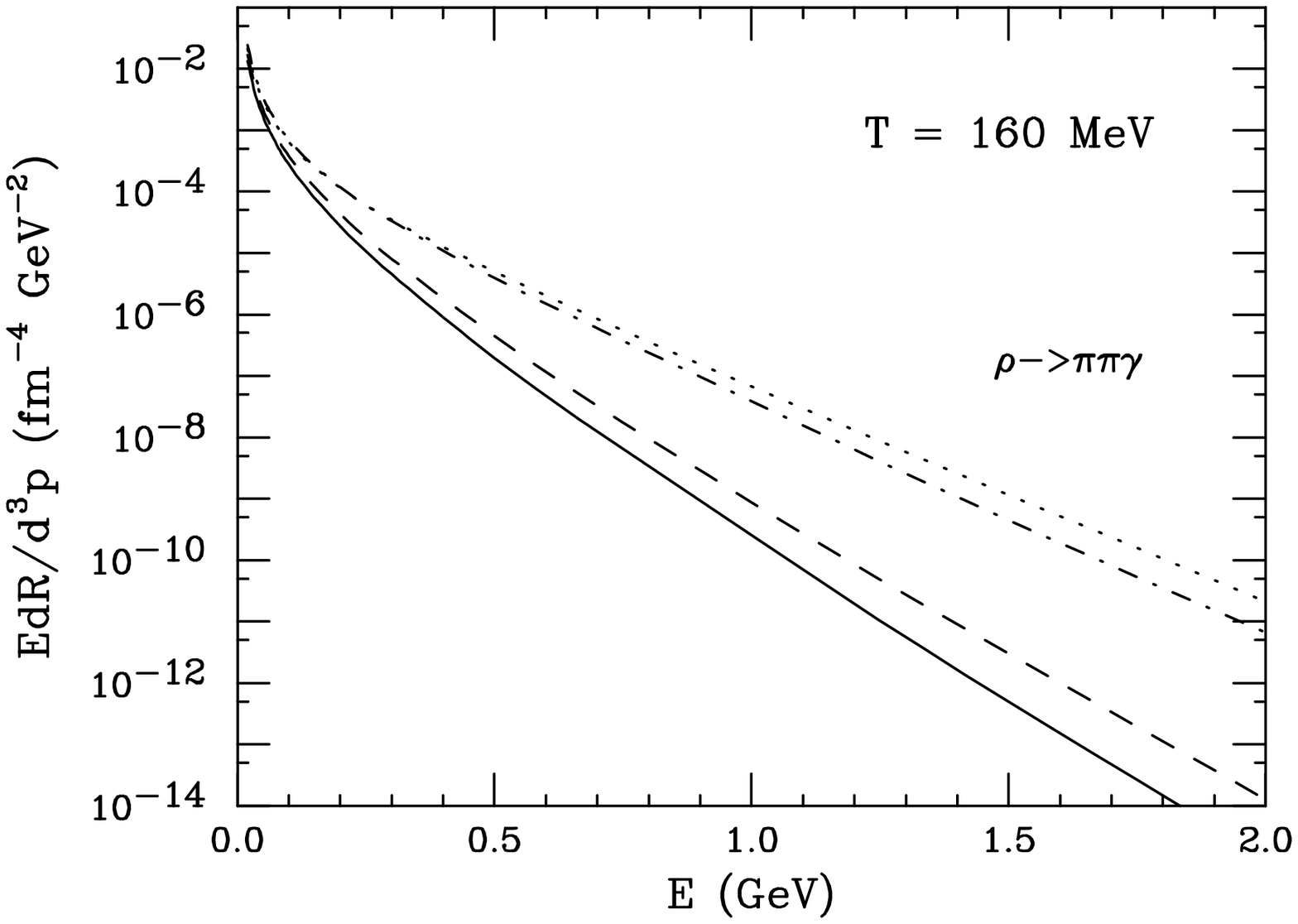,height=6cm,width=8cm}}
\caption{
Same as Fig.~\protect\ref{fig8} for the decay $\rho\,\ra\,\pi\,\pi\,\gamma$. 
}
\label{fig12}
\eef

\bef
\centerline{\psfig{figure=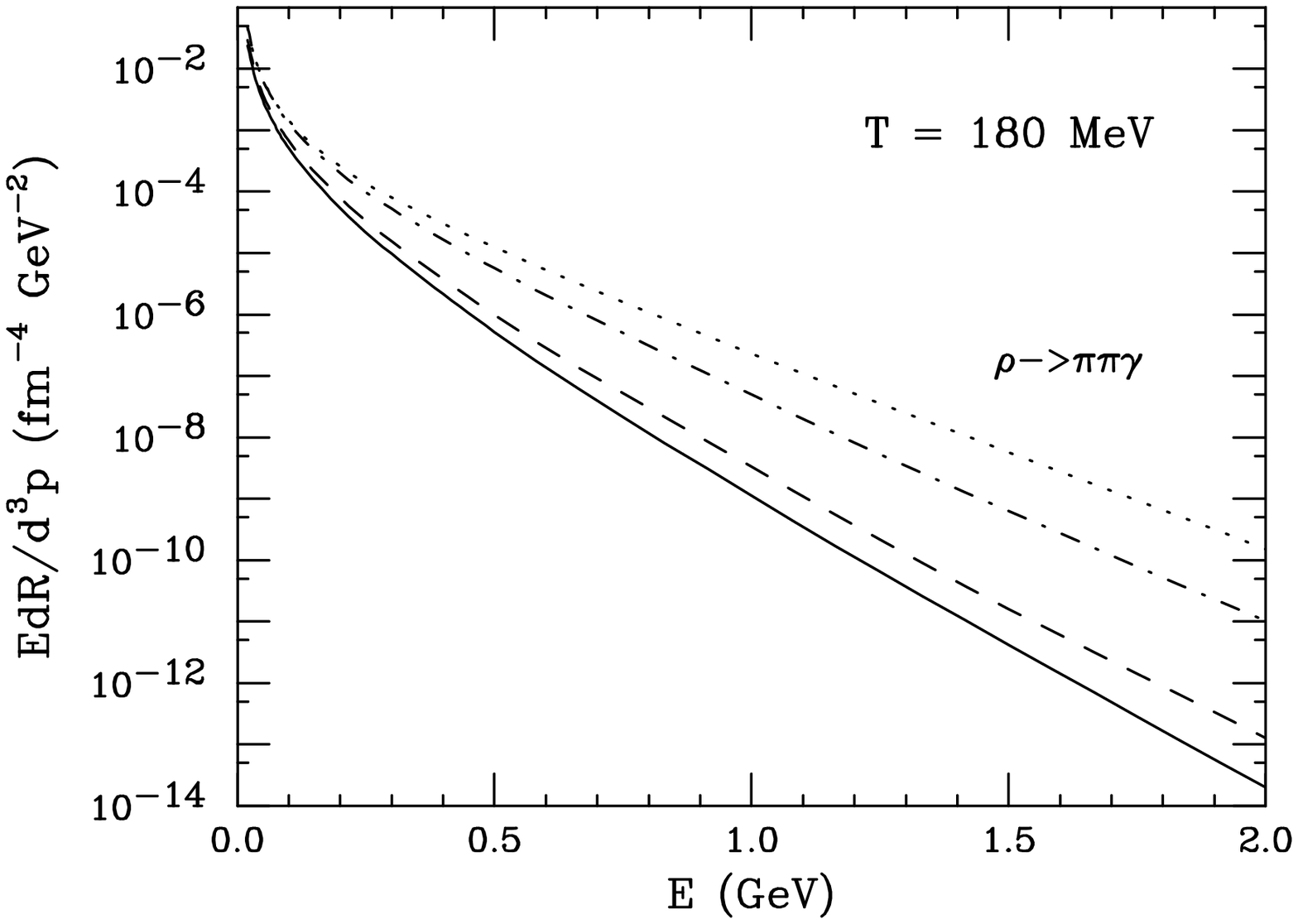,height=6cm,width=8cm}}
\caption{
Same as Fig.~\protect\ref{fig12} at $T$ = 180 MeV. 
}
\label{fig13}
\eef

Since the lifetimes of the  rho and omega mesons are comparable to the strong
interaction time scale it is necessary to consider their decay channels.
The possibility of detecting a high energy photon from the decay
$\rho\,\ra\,\pi\,\pi\,\gamma$ reduces as the mass of the rho decreases
with increasing baryon density. This is clearly observed from
 Fig.~(\ref{fig12}) at $T$ = 160 MeV. The corresponding results 
for $T$ = 180 MeV (Fig.~(\ref{fig13})) shows a similar nature.
In Figs.~(\ref{fig14}) and (\ref{fig15}) we display the photon spectra 
from omega decay for $T$ = 160 and 180 MeV respectively. The medium
effects in this case are rather small.

\bef
\centerline{\psfig{figure=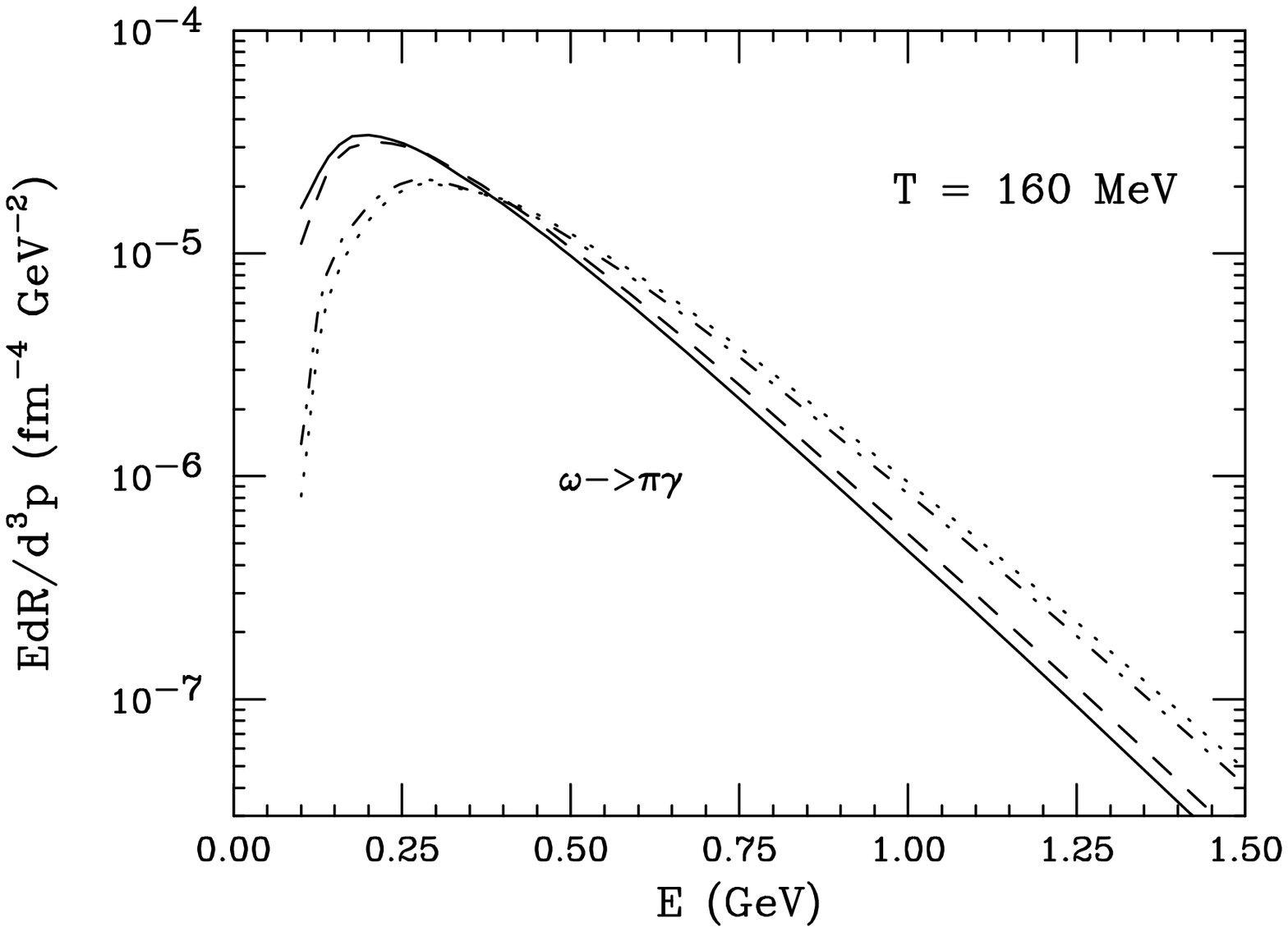,height=6cm,width=8cm}}
\caption{
Same as Fig.~\protect\ref{fig8} for the decay $\omega\,\ra\,\pi\,\gamma$. 
}
\label{fig14}
\eef

\bef
\centerline{\psfig{figure=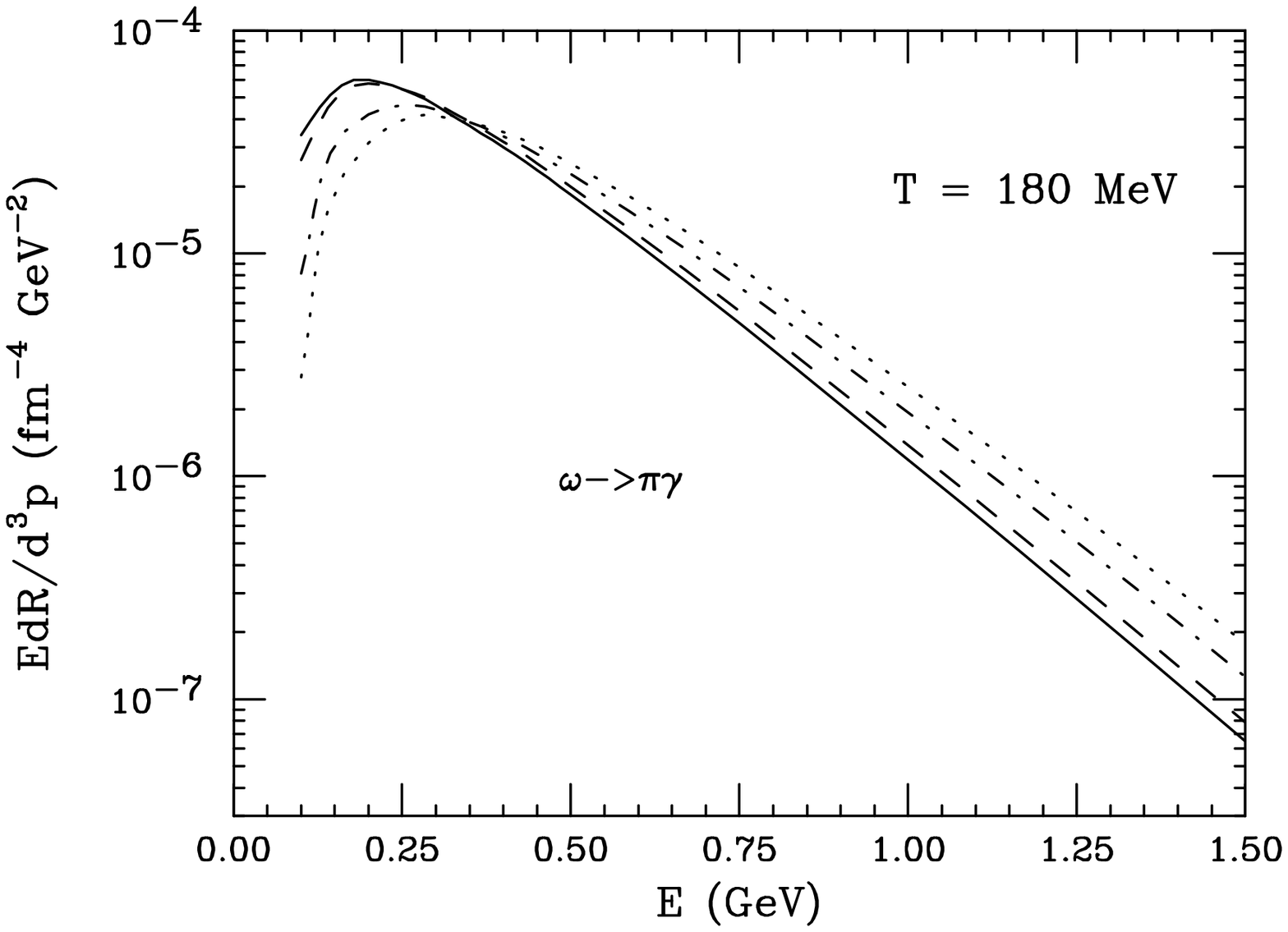,height=6cm,width=8cm}}
\caption{
Same as Fig.~\protect\ref{fig14} at $T$ = 180 MeV. 
}
\label{fig15}
\eef


Finally, the total photon emission rate at $T$=160 MeV is plotted in 
Fig.~(\ref{fig16})
as a function of the energy of the emitted photon for various values
of baryon density. At twice normal nuclear matter density the photon
yield is seen to be higher by a factor of $\sim$ 4 
compared to the case when the effects of the thermal bath
on the hadronic properties are neglected. This is true for almost the
entire energy range of the emitted photon under consideration. 
A qualitatively similar trend is observed at $T=$ 180 MeV as 
shown in Fig.~(\ref{fig17}). As mentioned earlier, the delta-hole
polarization effects on pion and its consequences on photon
spectra may be important which has been neglected in the present 
work.

    In order to accout for the fact that most of the photon producing
reactions which we have considered contain unstable particles (rho
and omega) in the external lines we have folded the emission rate with
the appropriate spectral function of the unstable particle as given in
eqs.~(11) and (12). The limits of the $s$ integration in the case of different
channel have been determined from kinematical considerations. 
The results for the reactions $\pi\,\pi\,\ra\,\rho\,\gamma$, $\pi\,\pi\,\ra\,
\pi\,\gamma$, $\rho\,\ra\,\pi\,\pi\,\gamma$ and $\omega\,\ra\,\pi\,\gamma$ have
been shown in Fig.~(\ref{fig18}). The difference caused by the inclusion of the
spectral function is observed to be $\sim$ a few percent.
This is reflected in the total photon spectra as shown in Fig.~(\ref{fig19}).

From the set of photon producing reactions considered in this
work it is clear that the rho and not the omega meson plays the dominant
role. Therefore the separation of the longitudinal and transverse 
mode in case of omega meson has negligible effect on the total 
photon emission rate. In our calculation the full energy momentum 
dependence of the self energy of rho enters through the propagator
(eq.~(\ref{deff})). However, we have observed that the quantity 
$k_0^2-{\bd k}^2$ along the dispersion curve 
(Fig.~(\ref{fig3})) remains almost constant ($\sim m_{\rho}^{\ast 2}$,
which is defined as $k_0^2$ at ${\bd k}^2=0$ of the mass hyperbola).
In other words in this particular case as far as the photon emission
rate is concerned a simple pole approximation of the rho propagator
at $k^2=m_{\rho}^2$ reproduces the results obtained with the full
propagator.

So far we have studied the effects of in-medium hadronic masses
and decay widths on the photon spectra, where, 
the latter is found to have negligible effects. However, the modification in the
rho decay width due to BE, arising from the induced emission of 
pions in the thermal medium, plays  a crucial role in  low mass 
lepton pair production. It has been shown
in Ref.~~\cite{weldon} that this effect arises naturally in a calculation 
based on finite temperature field theory.

For the sake of illustration,
we consider the effect of thermal nucleon loop on the lepton pair 
production from pion annihilation via (rho mediated) vector meson 
dominance. In the work of Li, Ko and Brown~\cite{li}, the observed
enhancement on the low mass dilepton yield was attributed solely to the
the matter induced mass modification of the
meson, while neglecting the BE effect. 
In Fig.~(\ref{fig20}) the dilepton emission rate is plotted 
as a function of invariant mass at a temperature  $T$=180 MeV. 
It is clear from the figure that the inclusion of BE effects 
reduces the yield by increasing the decay width. Quantitatively,
a suppression by a factor of 3 is observed at the invariant mass,
$M=m_{\rho}^{\ast}$.

\bef
\centerline{\psfig{figure=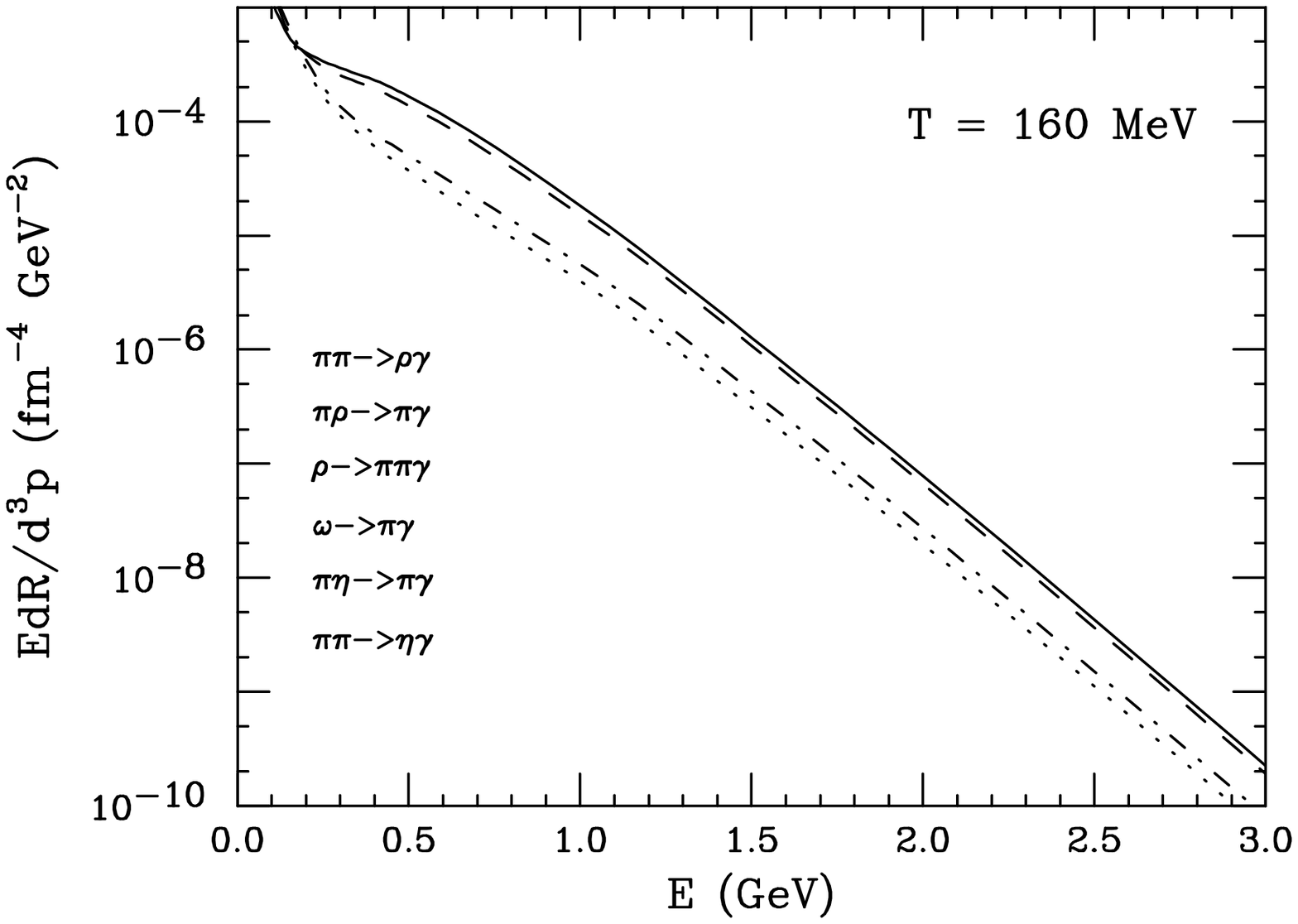,height=6cm,width=8cm}}
\caption{
Total photon emission rates as a function of energy of the emitted photon
at $T$ = 160 MeV. Solid and dashed lines show
the results for $\rho/\rho_0$ = 2 and 1 respectively.
The dotdashed line represents only finite temperature effects and dotted
line shows the result when no medium effect is considered.
}
\label{fig16}
\eef

\bef
\centerline{\psfig{figure=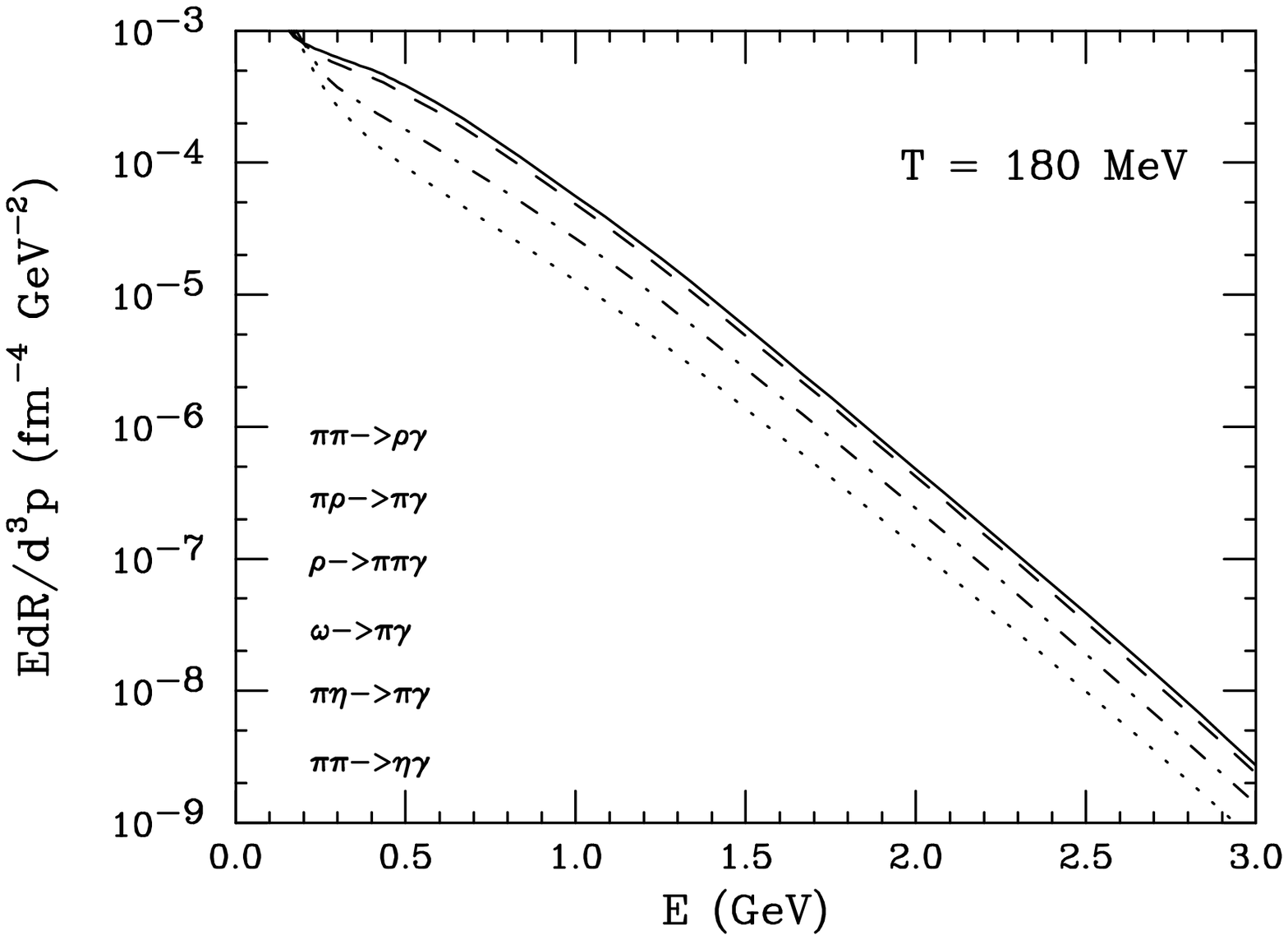,height=6cm,width=8cm}}
\caption{
Same as Fig.~\protect\ref{fig16} at $T$ = 180 MeV. 
}
\label{fig17}
\eef
\bef
\centerline{\psfig{figure=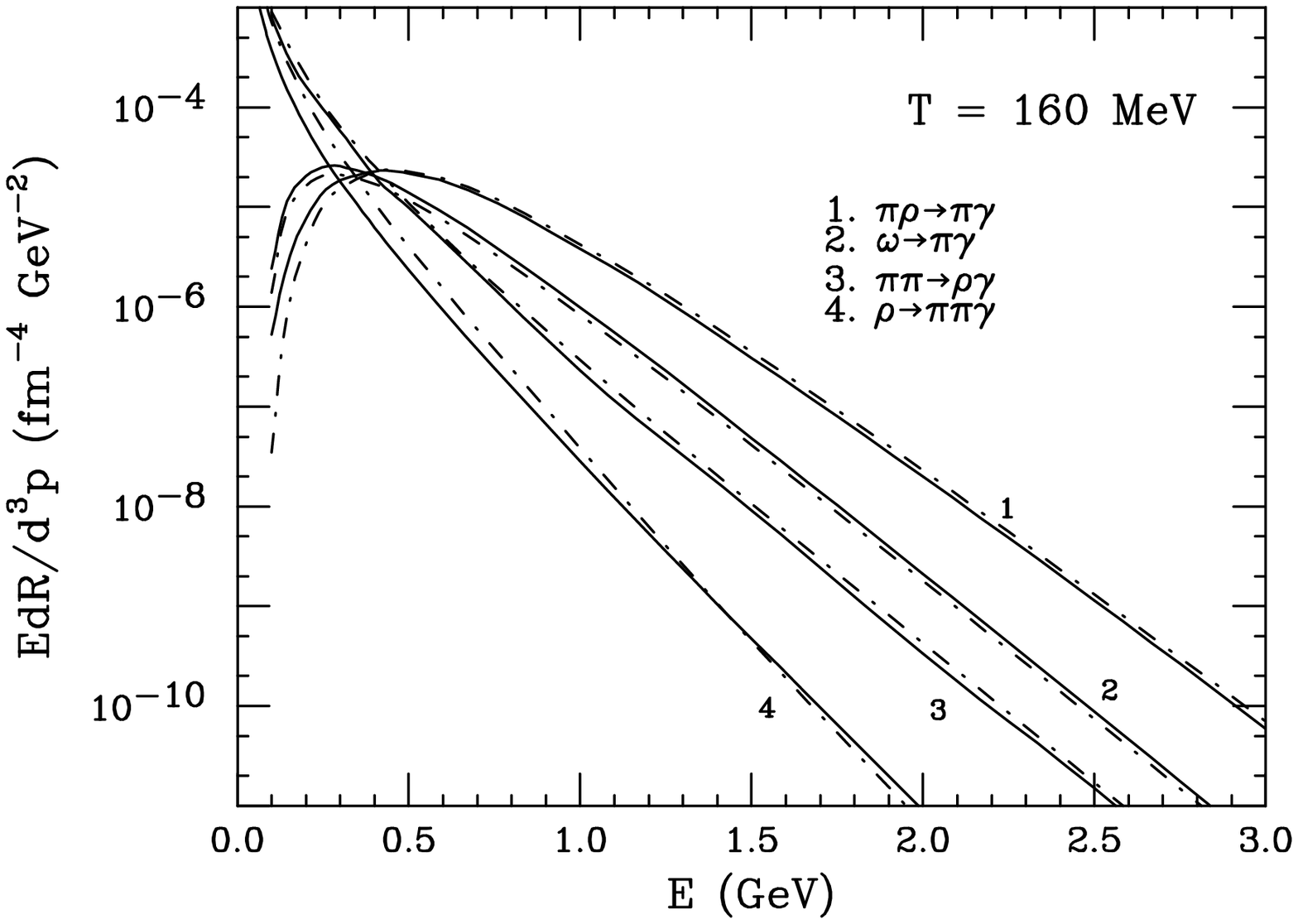,height=6cm,width=8cm}}
\caption{Effect of spectral function of vector mesons on photon emission
rates at $T$ = 160 MeV and zero baryon density. Solid (dotdashed) line shows
results with (without) the inclusion of spectral function.
}
\label{fig18}
\eef

\bef
\centerline{\psfig{figure=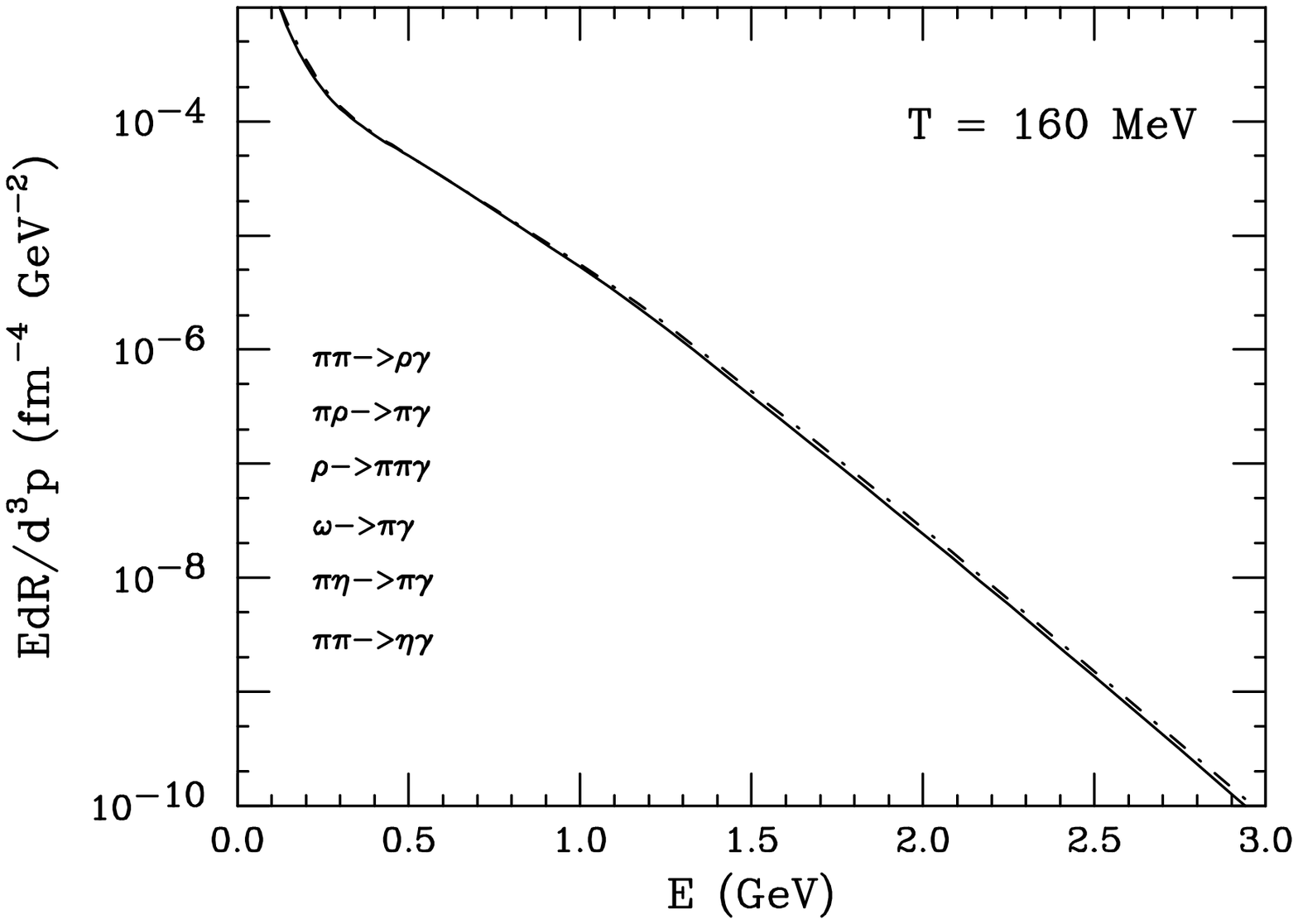,height=6cm,width=8cm}}
\caption{Total photon emission
rate at $T$ = 160 MeV and zero baryon density.
}
\label{fig19}
\eef

\bef
\centerline{\psfig{figure=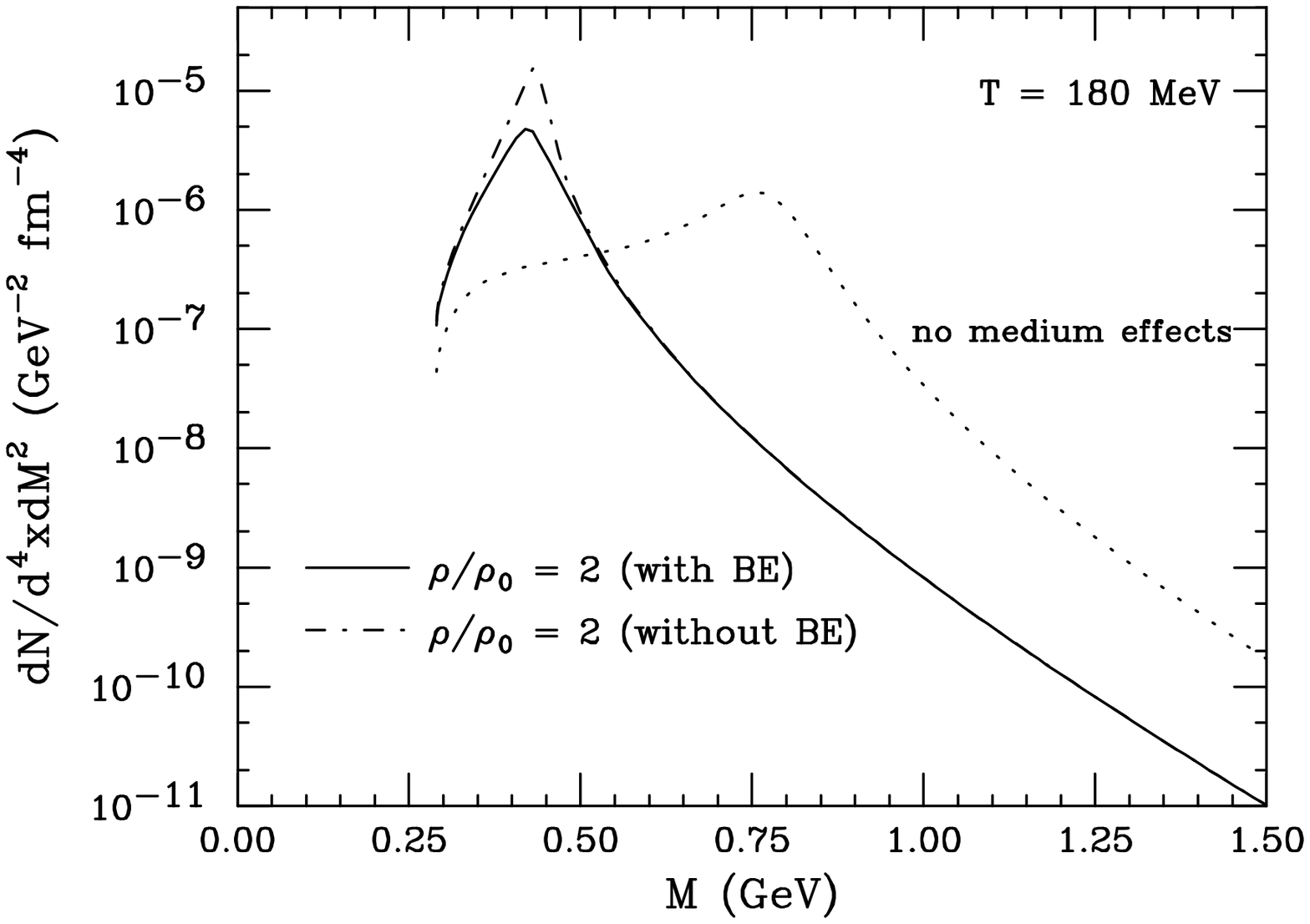,height=6cm,width=8cm}}
\caption{
Dilepton emission rates as a function of invariant mass 
at $T$ = 180 MeV and twice nuclear matter density. Solid and dotdashed 
lines show the results with and without BE effects respectively. The dotted
line shows the result when no medium effect is considered.
}
\label{fig20}
\eef

\section*{V. Summary and Discussions}

In this work we have studied the variation of effective masses
and decay widths of nucleons and vector mesons at non-zero
temperature and baryon density within the framework of an effective
Lagrangian approach. It is seen that the nucleon mass for non-zero
baryon density increases slightly with temperature before dropping.
We have seen that the effect of the vacuum fluctuation corrections
to MFT value of nucleon mass is substantial at higher values of
the baryon density. However, the consequence of such an 
effect on the vector meson mass is minimal. 

We have also estimated the photon emission rate from a hot
and dense hadronic matter. The inclusion of in-medium properties
of hadrons enhances the emission rate.
Although the effect of the decay width of the rho meson
on photon spectra is small, it affects the dilepton
emission rate quite substantially in the low invariant mass region (see
{\it e.g.} Ref.~\cite{Csong}).
An exhaustive work taking into account all the important
dilepton producing channels will be published shortly.

So far we have talked about the static (fixed temperature) 
emission rate of photons and dileptons. The calculated
spectra has to be folded with the space time evolution dynamics.
This requires a self consistent calculation of the equation of state,
which is a fundamental ingredient to any hydrodynamical calculation. 
Because of the substantial change in hadronic properties at 
finite temperature and density a simple Bjorken-like~\cite{bj}
hydrodynamical model appears to be grossly inadequate. 
A rigorous calculation addressing these issues is in progress.

\noindent {\bf Acknowledgement:}\\
We are grateful to Professor Binayak Dutta-Roy
for many useful discussions and also to the referee(s) whose
comments were very helpful.

\appendix
\section{}
The polarisation tensor of a vector (omega or rho) meson due to $VNN$ 
interaction is obtained
from the Lagrangian described by eq.~(\ref{lag1}) as
\be
\Pi^{\mu \nu} = -2ig_{VNN}^2\,\int\,\frac{d^4p}{(2\pi)^4}\,{\s{Tr}}\left[
\frac{}{}\Gamma^{\mu}(k)G(p)\Gamma^{\nu}(-k)G(p+k)\right]
\ee
where 
\be
\Gamma^{\mu}(k) = \gamma^{\mu}+i\frac{\kappa_V}{2M}\sigma^{\mu \alpha}
k_{\alpha} 
\ee
In a hot and dense medium moving with four velocity $u^{\mu}$ the fermion
propagator is given by~\cite{weldon1}
\ba
G(p)&=&  (\gamma^{\mu}{p\prime}_{\mu}+M^{\ast})\left[\frac{1}{{p\prime}^2-
M^{\ast 2}
+i\epsilon}+2\pi i\frac{\delta({p\prime}^2-M^{\ast 2})}{e^{|p\cdot u|}+1}
\right]\nunu\\
&=&G_F(p)+G_D(p)
\ea
with
${p\prime}^{\nu} = p^{\nu}+\mu^{\ast}u^{\nu}$ and $|p\cdot u| = 
\sqrt{|\bd p|^2+M^{\ast 2}}-\mu^{\ast}$.

The vacuum part of the vector meson self energy due to $N\bar N$ polarization
in the modified Dirac sea is given by
\be
\Pi_{\mu \nu}^F = -2ig_{VNN}^2\,\int\,\frac{d^4p}{(2\pi)^4}\,{\s{Tr}}\left[
\frac{}{}\Gamma^{\mu}(k)G_F(p)\Gamma^{\nu}(-k)G_F(p+k)\right]
\ee
with
\ba
\Pi^F(k^2)&=&\frac{1}{3}{\s{Re}}(\Pi^F)_{\mu}^{\mu}\nunu\\
&=&-\frac{g_{VNN}^2}{\pi^2}
k^2\,\left[I_1+M^{\ast}\frac{\kappa_V}{2M}I_2+\frac{1}{2}\,(\frac{\kappa_V}
{2M})^2\,(k^2I_1+M^{\ast 2}I_2)\right]
\label{pik2}
\ea
where,
\be
I_1=\int_{0}^{1}\,dz\,z(1-z)\,\ln\left[\frac{M^{\ast 2}-k^2\,z(1-z)}
{M^2-k^2\,z(1-z)}\right],
\ee
\be
I_2=\int_{0}^{1}\,dz\,\ln\left[\frac{M^{\ast 2}-k^2\,z(1-z)}
{M^2-k^2\,z(1-z)}\right].
\ee
Here, it is important to note that we have used the in-medium mass
obtained by solving Eqs.~(\ref{nmass}) and (\ref{bden}) simultaneously. 
The  renormalization scheme employed by Hatsuda et al.~\cite{hatsuda}.
has been adopted here.

In a hot and dense medium, because of Lorentz invariance and current
conservation the general structure of the polarisation tensor is of the form
\be
\Pi^{\mu \nu} = \Pi_T(k_0,|\bd k|)A^{\mu \nu}+\Pi_L(k_0,|\bd k|)B^{\mu \nu}
\ee
where the two Lorentz invariant functions $\Pi_T$ and $\Pi_L$ are 
obtained by contraction:
\ba
\Pi_L&=&-\frac{k^2}{|\bd k|^2}u^{\mu}u^{\nu}\Pi_{\mu \nu}\nonumber\\
\Pi_T&=&\frac{1}{2}(\Pi_{\mu}^{\mu}-\Pi_L)
\ea

In the case of the vector meson interacting with real particle-hole
excitations in the nuclear medium these are given by
\ba
\Pi_{\mu \nu}^D &=& -2ig_{VNN}^2\,\int\,\frac{d^4p}{(2\pi)^4}\,{\s{Tr}}\left[
\frac{}{}\Gamma^{\mu}(k)G_F(p)\Gamma^{\nu}(-k)G_D(p+k)+(F\leftrightarrow D)
\right]\nunu\\
&=&(\Pi_v^D+\Pi_{vt}^D+\Pi_t^D)_{\mu \nu}
\ea
with
\ba
(\Pi_v^D)_{\mu}^{\mu} &=& \frac{g_{VNN}^2}
{2\pi^2}\,\frac{1}
{|\bd k|}\,\int\,\frac{pdp}{\omega_p}\,\left[(k^2+2M^{\ast 2})
\ln\left\{
\frac{(k^2+2|\bd p||\bd k|)^2-4k_0^2\omega_p^2}{(k^2-2|\bd p||\bd k|)^2-
4k_0^2\omega_p^2}\right\}\right.\nonumber\\
&&-\left.\frac{}{}8|\bd p||\bd k|\right]
\left[\frac{}{}n_B(\mu^{\ast},T)+{\bar n}_B(\mu^{\ast},T)\right]
\ea
\ba
(\Pi_{vt}^D)_{\mu}^{\mu} &=& \frac{3g_{VNN}^2}{\pi^2}\,
M^{\ast}\left(\frac{\kappa_V}{2M}\right)\,\frac{k^2}{|\bd k|}\,\int\,
\frac{pdp}{\omega_p}\,
\ln\left\{
\frac{(k^2+2|\bd p||\bd k|)^2-4k_0^2\omega_p^2}{(k^2-2|\bd p||\bd k|)^2-
4k_0^2\omega_p^2}\right\}\nonumber\\
&&\times
\left[\frac{}{}n_B(\mu^{\ast},T)+{\bar n}_B(\mu^{\ast},T)\right]
\ea
\ba
(\Pi_t^D)_{\mu}^{\mu}&=& \frac{g_{VNN}^2}{4\pi^2}\,
\left(\frac{\kappa_V}{2M}\right)^2\frac{k^2}
{|\bd k|}\,\int\,\frac{pdp}{\omega_p}\,\left[\frac{}{}(k^2+8M^{\ast 2})
\right.\nonumber\\
&&\times\left.
\ln\left\{
\frac{(k^2+2|\bd p||\bd k|)^2-4k_0^2\omega_p^2}{(k^2-2|\bd p||\bd k|)^2-
4k_0^2\omega_p^2}\right\}
-4|\bd p||\bd k|\right]\nonumber\\
&&\times\left[\frac{}{}n_B(\mu^{\ast},T)+{\bar n}_B(\mu^{\ast},T)\right]
\ea
The longitudinal component of the polarisation tensor is given by
\be
\Pi_L^D=\Pi_L^{D,v}+\Pi_L^{D,vt}+\Pi_L^{D,t}
\ee
with
\ba
\Pi_L^{D,v}&=&-\frac{g_{VNN}^2}{4\pi^2}\,
\frac{k^2}{|\bd k|^3}
\int\,\frac{pdp}{\omega_p}\,\left[\frac{}{}\{(k_0-2\omega_p)^2-|\bd k|^2\}
\ln{\frac{k^2-2k_0\omega_p+2|\bd p||\bd k|}{k^2-2k_0\omega_p-2|\bd p||\bd k|}}
\right.\nonumber\\
&&+\left.\{(k_0+2\omega_p)^2-|\bd k|^2\}
\ln{\frac{k^2+2k_0\omega_p+2|\bd p||\bd k|}{k^2+2k_0\omega_p-2|\bd p||\bd k|}}
-8|\bd p||\bd k|\right]
\nonumber\\
&&\times\left[\frac{}{}n_B(\mu^{\ast},T)+{\bar n}_B(\mu^{\ast},T)\right]
\ea
\ba
\Pi_L^{D,vt} &=&\frac{g_{VNN}^2}{\pi^2}\,
M^{\ast}\left(\frac{\kappa_V}{2M}\right)
\frac{k^2}{|\bd k|}
\int\,\frac{pdp}{\omega_p}\,
\ln\left\{
\frac{(k^2+2|\bd p||\bd k|)^2-4k_0^2\omega_p^2}{(k^2-2|\bd p||\bd k|)^2-
4k_0^2\omega_p^2}\right\}\nonumber\\
&&\times\left[\frac{}{}n_B(\mu^{\ast},T)+{\bar n}_B(\mu^{\ast},T)\right]
\ea

\ba
\Pi_L^{D,t}&=&-\frac{g_{VNN}^2}{2\pi^2}\,
\left(\frac{\kappa_V}{2M}\right)^2
\frac{k^2}{|\bd k|}
\int\,\frac{pdp}{\omega_p}\,
\left[\left\{2|\bd p|^2-\frac{k^2}{2}-\frac{(k^2-2k_0\omega_p)^2}{2|\bd k|^2}
\right\}
\right.\nonumber\\
&&\times\left.\ln{\frac{k^2-2k_0\omega_p+2|\bd p||\bd k|}
{k^2-2k_0\omega_p-2|\bd p||\bd k|}}
+\left\{2|\bd p|^2-k^2-\frac{(k^2+2k_0\omega_p)^2}{|\bd k|^2}\right\}
\right.\nonumber\\
&&\times\left.\ln{\frac{k^2+2k_0\omega_p+2|\bd p||\bd k|}
{k^2+2k_0\omega_p-2|\bd p||\bd k|}}
-\frac{4|\bd p|k_0^2}{|\bd k|}\right]\nonumber\\
&&\times\left[\frac{}{}n_B(\mu^{\ast},T)+{\bar n}_B(\mu^{\ast},T)\right]
\ea
The dispersion relation for the longitudinal (transverse) mode now reads 
\be
k_0^2-|\bd k|^2-m_V^2+{\s {Re}}\Pi_{L(T)}^D(k_0,{\bd k})+{\s{Re}}
\Pi^F(k^2) = 0
\label{disp}
\ee

Usually the physical mass $(m_V^{\ast})$ is defined as the lowest
zero of the above equation in the limit ${\bd k}\ra 0$. 
In this limit
$\Pi_T^D = \Pi_L^D = \Pi^D$, and we have,
\be
\frac{1}{3}\Pi_{\mu}^{\mu}=\Pi= \Pi^D+\Pi^F
\ee
where
\be
\Pi^D(\omega,{\bd k}\ra 0) =  
-\frac{4g_{VNN}^2}{\pi^2}\,\int\,p^2\,dp\,F(|\bd p|,M^{\ast})\,
[\frac{}{}n_B(\mu^{\ast},T)+{\bar n}_B(\mu^{\ast},T)]
\ee
with
\ba
F(|\bd p|,M^{\ast})&=&\frac{1}{\omega_p(4\omega_p^2-\omega^2)}
\,\left[\frac{2}{3}(2|\bd p|^2+3M^{\ast 2})+\omega^2\left\{2M^{\ast}
(\frac{\kappa_V}{2M})\right.\right.\nonumber\\
&&+\left.\left.\,\frac{2}{3}(\frac{\kappa_V}{2M})^2(|\bd p|^2
+3M^{\ast 2})\right\}\right]
\ea
where $\omega_p^2={\bd p}^2+M^{\ast 2}$.

The effective mass of the vector meson is then obtained by solving the
equation:

\be
k_0^2 - m_V^2 + {\mathrm {Re}}\Pi = 0
\label{mass}
\ee

One finds reference to two other kinds of masses in the literature. 
The invariant mass is defined
as the lowest order zero of eq.(\ref{disp}) with $\Pi^D$ neglected. 
Again, the screening mass of a vector
meson is obtained from the pure imaginary zero of the quantity
on the left hand side of the same equation with $k_0=0$.

The imaginary part of the self energy of a particle 
is related to the probability of its survival in a  
medium at finite temperature. For a rho meson propagating
with energy $\omega > 0$ and three momentum $\bd k$ this is given by  
\be
\Gamma(\omega) = \frac{g_{\rpp}^2}{48\pi}\,W^3(s)\,\frac{s}{\omega}\,
\left[1+\frac{2T}{W(s)\sqrt{\omega^2-s}}\,
\ln\left\{\frac{1-\exp[-\frac{\beta}{2}(\omega+W(s)\sqrt{\omega^2-s})]}
{1-\exp[-\frac{\beta}{2}(\omega-W(s)\sqrt{\omega^2-s})]}\right\}
\right]
\ee
where $s = k^2 = \omega^2-{\bd k}^2$ and $W(s) = \sqrt{1-4m_{\pi}^2/s}$.
In the limit $|\bd k| \ra 0$ , the above expression reduces to 
the in-medium decay width  and is given by
\be
\Gamma_{\rho\,\rightarrow\,\pi\,\pi} = \frac{g_{\rho\,\pi\,\pi}^2}{48\pi}\,
\omega\,W^3(\omega)
\left[\left(1+n(\frac{\omega}{2})\right)\,\left(1+n(\frac{
\omega}{2})\right)-n(\frac{\omega}{2})n(\frac{\omega}{2})
\right]
\label{width}
\ee
where $\omega=m_{\rho}^\ast$ is the in-medium mass of the rho due to
$N-N$ interaction.

\section{}
\bef
\centerline{\psfig{figure=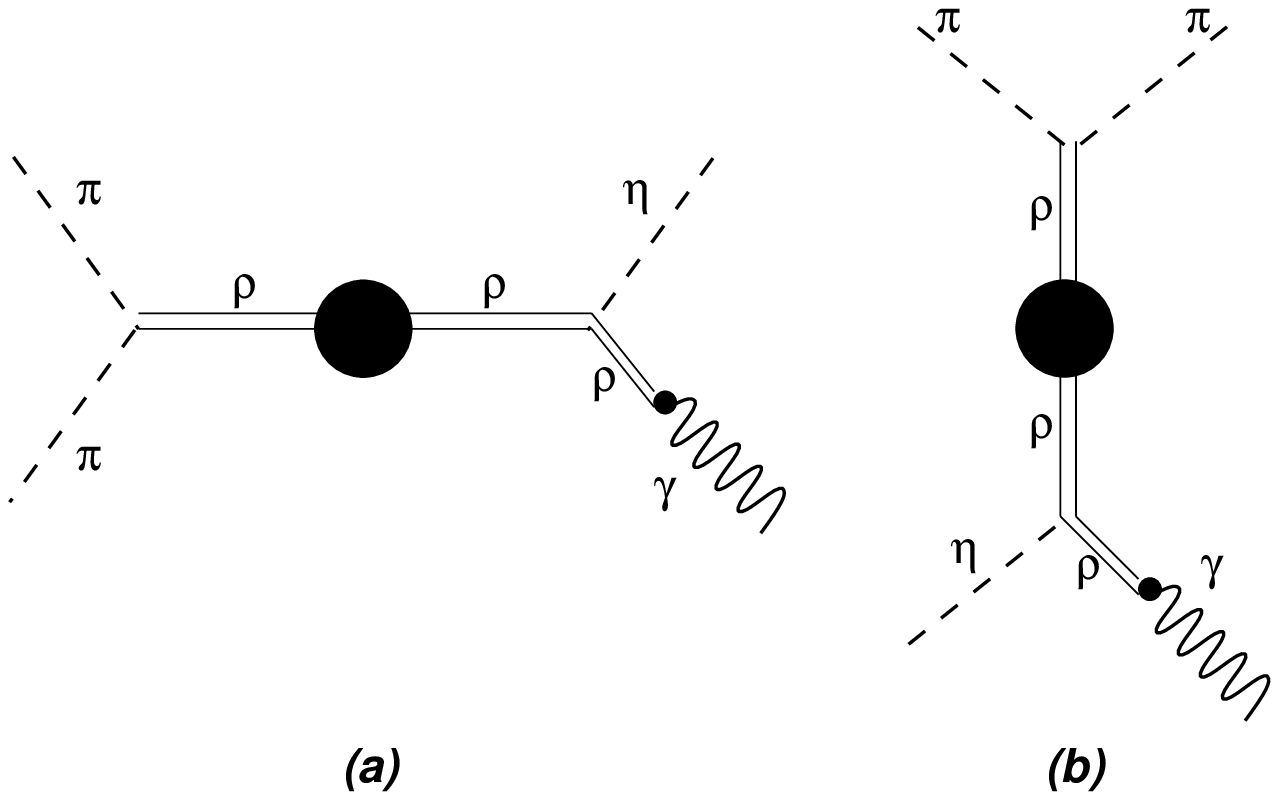,height=6cm,width=8cm}}
\caption{
Feynman diagrams for $\pi\,\pi\,\ra\,\eta\,\gamma$ and $\pi\,\eta\,\ra\,
\pi\,\gamma$.
}
\label{fig21}
\eef

The $\rho \pi \eta$ interaction is described by Eq.~(\ref{etaro}).
The invariant amplitude for the reaction $\pi^+(p_1)+\pi^-(p_2)\,
\ra\,\eta(p_3)+\gamma(p_4)$ is given by
\ba
|{\cal M}_a|^2&=&\frac{4\pi\,\alpha\,g_{\rho \rho \eta}^2}
{m_{\eta}^2[(s-m_{\rho}^2)^2+m_{\rho}^2\Gamma_{\rho}^2]}\,
\left[\frac{}{}s(u-m_{\pi}^2)(t-m_{\pi}^2)-m_{\pi}^2(s-m_{\eta}^2)^2\right]
\ea
The Feynman diagram for the above process is shown in Fig.~(\ref{fig21}a).
For the reaction $\pi^{\pm}(p_1)+\eta(p_2)\, \ra\,\pi^{\pm}(p_3)+\gamma(p_4)$
(see Fig.~(\ref{fig21}b)),
\ba
|{\cal M}_b|^2&=&\frac{4\pi\,\alpha\,g_{\rho \rho \eta}^2}
{m_{\eta}^2[(t-m_{\rho}^2)^2+m_{\rho}^2\Gamma_{\rho}^2]}\,
\left[\frac{}{}t(u-m_{\pi}^2)(s-m_{\pi}^2)-m_{\pi}^2(t-m_{\eta}^2)^2\right]
\ea
The coupling constant $g_{\rho \rho \eta}$ is evaluated from the following 
relations:
\ba
\Gamma(\rho\,\ra\,\eta\,\gamma)&=&\frac{(m_{\rho}^2-m_{\eta}^2)^3}
{96\pi\,m_{\eta}^2\,m_{\rho}^3}\,g_{\eta \gamma \rho}^2\nonumber\\
g_{\eta \gamma \rho}&=&\frac{e\,g_{\rho \rho \eta}}{g_{\rho \pi \pi}}
\nonumber\\
\frac{g_{\rho \pi \pi}^2}{4\pi}&=&2.9
\ea
and $\Gamma(\rho\,\ra\,\eta\,\gamma) = (57 {\pm} 10.5)$ keV.


\begin{thebibliography}{99}

\bibitem{ukawa} A. Ukawa, in Quark Matter'97, Tsukuba, Japan (Proc.
to be published in Nucl. Phys. {\bf A}).

\bibitem{hwa} R. C. Hwa (ed.), Quark-Gluon Plasma, World Scientific,
Singapore 1990.

\bibitem{janepr} J. Alam, S. Raha, and B. Sinha, Phys. Rep. {\bf 273}
243 (1996).

\bibitem{pisarski} R. D. Pisarski, hep-ph/9503330.

\bibitem{meissner} U. Meissner, Phys. Rep. {\bf 161} 213 (1988).

\bibitem{furn} R. J. Furnstahl and T. Hatsuda, Phys. Rev. {\bf D42}
 1744 (1990); C. Adami, T. Hatsuda and I. Zahed, Phys. Rev. {\bf D43}
921 (1991); T. Hatsuda, Nucl. Phys. {\bf A544} 27c (1992).

\bibitem{brown} G. E. Brown, Nucl. Phys. {\bf A522} 397c (1991).

\bibitem{rho} G. E. Brown and M. Rho, Phys. Rev. Lett. {\bf 66} 2720 (1991).

\bibitem{abhijit} A. Bhattacharyya, J. Alam, S. Raha and B. Sinha,
Int. J. Mod. Phys. {\bf A12} 5639 (1997).

\bibitem{vol16} B. D. Serot and J. D. Walecka, Advances in Nuclear Physics
Vol. 16 Plenum Press, New York 1986. 

\bibitem{chin} S. A. Chin, Ann. Phys. {\bf 108} 301 (1977).

\bibitem{npa1} S. Sarkar, J. Alam, P. Roy, A. K. Dutt-Majumder, B. Dutta-Roy
and B. Sinha, Nucl. Phys. {\bf A634} 206 (1998).

\bibitem{qm96} See for example, Nucl. Phys. {\bf A610} (1996).

\bibitem{weldonann} H. A. Weldon, Ann. Phys. {bf 228} 43 (1993).

\bibitem{prc99} J. Alam, S. Sarkar, P. Roy, B. Dutta-Roy and B. Sinha
Phys. Rev. {\bf C59} 905 (1999).

\bibitem{bellac} M. Le Bellac, Thermal Field Theory, (Cambridge University
Press, N. Y., 1996)

\bibitem{abrikosov} A. A. Abrikosov, L. P. Gor'kov and I. E. Dzyaloshinski,
Method of Quantum Field Theory in Statistical Physics, 
(Prentice Hall, Engelwood Cliffs, N. J. 1963).

\bibitem{zubarev} D. N. Zubarev, Sov. Phys. Uspekhi, {\bf 3} 69 (1960).

\bibitem{rapp} R. Rapp, G. Chanfray and J. Wambach, Nucl. Phys. {\bf A617}
472 (1997).

\bibitem{sakurai} J. J. Sakurai, Currents and Mesons, The University
of Chicago Press, Chicago, 1969.

\bibitem{jean} H. C. Jean, J. Piekarewicz and A. G. Williams,
Phys. Rev. {\bf C49}, 1981(1994).

\bibitem{li} G. Q. Li, C. M. Ko and G. E. Brown, Nucl. Phys. {\bf A606}
568 (1996).

\bibitem{kapusta} J. I. Kapusta, Finite Temperature Field Theory,
Cambridge University Press, 1993.

\bibitem{hatsuda1} T. Hatsuda, private communication.

\bibitem{gale} C. Gale and J. I. Kapusta, Nucl. Phys. {\bf B357} 65 (1991).

\bibitem{klingl} F. Klingl, N. Kaiser and W. Weise, Nucl. Phys. {\bf A624} 
527 (1996).

\bibitem{asakawa} M. Asakawa, C. M. Ko, P. Levai and X. J. Qiu
Phys. Rev. {\bf C46} R1159 (1992).

\bibitem{friman} M. Herrmann, B. L. Friman and 
W. N\"orenberg, Nucl. Phys. {\bf A560} 411 (1993).

\bibitem{chanfray} G. Chanfray and P. Shuck, Nucl. Phys. {\bf A545} 271c (1992).

\bibitem{abhee} A. K. Dutt-Mazumder, J. Alam, B. Dutta-Roy and Bikash
Sinha, Phys. Lett. {\bf B378} 35 (1996).  

\bibitem{joe} J. Kapusta, P. Lichard, and D. Seibert, Phys. Rev.
{\bf D44} 2774 (1991).

\bibitem{weldon} H. A. Weldon, Phys. Rev. {\bf D28} 2007 (1983).

\bibitem{Csong} C. Song and C. M. Ko, Phys. Rev. {\bf C53} 2371 (1996).

\bibitem{bj} J. D. Bjorken, Phys. Rev. {\bf D27} 140 (1983).

\bibitem{weldon1} H. A. Weldon, Phys. Rev. {\bf D26} 1394 (1982).

\bibitem{hatsuda} T. Hatsuda, H. Shiomi and H. Kuwabara, Prog. Th. Phys.
{\bf 95} 1009 (1996).

\end{thebibliography}
\end{document}